\documentclass[manuscript, screen, nonacm]{acmart}
\AtBeginDocument{%
  }

\newcommand{\userquote}[1]{\textit{``#1''}}

\usepackage{booktabs}  % For better looking tables
\usepackage{amsmath}  % For math symbols
\usepackage{tabularx}
\usepackage{multirow}
\usepackage{arydshln}
\newcolumntype{R}{>{\raggedleft\arraybackslash}X} % Right 
\usepackage{ragged2e}
\newcolumntype{Y}{>{\RaggedRight\arraybackslash}X}

\usepackage{graphicx}
\usepackage{subcaption}
\usepackage{adjustbox}
\usepackage{multirow} 
\usepackage{ulem}
\usepackage{svg}

\usepackage{circledsteps}

\usepackage{cleveref}
\usepackage{listings}
\usepackage{pifont}
\usepackage{pifont}

\usepackage{csquotes}

\usepackage[most]{tcolorbox}

\usepackage{xcolor}
\definecolor{grey}{RGB}{128,128,128}
\definecolor{neonfuchsia}{rgb}{1.0, 0.25, 0.39}
\definecolor{darkgreen}{rgb}{0.0, 0.5, 0.0} % Define dark green color
\definecolor{darkcyan}{HTML}{00717f}

\tcbuselibrary{skins,breakable,raster}
\definecolor{DiffAdd}{HTML}{1F77B4}    % blue
\definecolor{DiffAgent}{HTML}{8A2BE2}  % purple

\newcommand{\diffctx}[1]{%
  \par\medskip\noindent\textbf{#1}\par
  \vspace{2pt}\hrule height 0.4pt\medskip
}

\newcommand{\diffnochg}[1]{#1}
\newcommand{\diffadd}[1]{\textcolor{DiffAdd}{\textbf{[+]}~#1}}
\newcommand{\diffagent}[1]{\textcolor{DiffAgent}{\textbf{[COPIED]}~#1}}

\setcopyright{acmlicensed}
\copyrightyear{2018}
\acmYear{2018}
\acmDOI{XXXXXXX.XXXXXXX}
\acmConference[Conference acronym 'XX]{Make sure to enter the correct
  conference title from your rights confirmation email}{June 03--05,
  2018}{Woodstock, NY}
\newcommand{\systemName}{\textsc{Perspectra}}
\acmISBN{978-1-4503-XXXX-X/2018/06}

\begin{document}

\title{\includegraphics[height=.8\baselineskip]{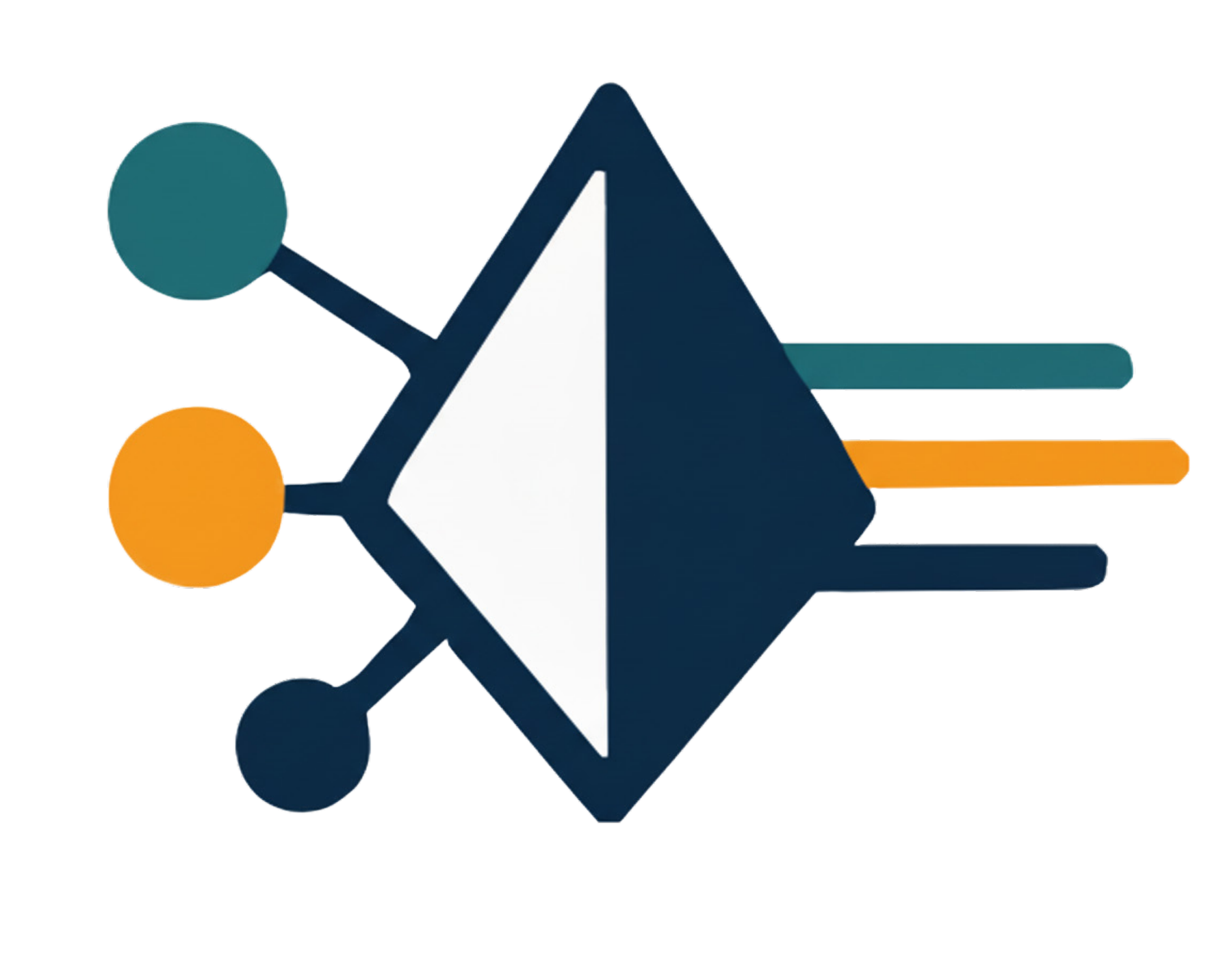}\systemName: Choosing Your Experts Enhances Critical Thinking in Multi-Agent Research Ideation}

%% Authors and affiliations
\author{Yiren Liu}
\affiliation{%
  \institution{Informatics, University of Illinois Urbana-Champaign}
  \city{Champaign}
  \state{Illinois}
  \country{USA}}
\email{yirenl2@illinois.edu}

\author{Viraj Shah}
\affiliation{%
  \institution{Siebel School of Computing and Data Science, University of Illinois Urbana-Champaign}
  \city{New York City}
  \state{New York}
  \country{USA}}
\email{virajns2@illinois.edu}

\author{Sangho Suh}
\affiliation{%
  \institution{Department of Computer Science, University of Toronto}
  \city{Toronto}
  \state{Ontario}
  \country{Canada}}

\author{Pao Siangliulue}
\affiliation{%
  \institution{Allen Institute for AI}
  \city{Seattle}
  \state{Washington}
  \country{USA}}

\author{Tal August}
\affiliation{%
  \institution{Siebel School of Computing and Data Science, University of Illinois Urbana-Champaign}
  \city{Urbana}
  \state{Illinois}
  \country{USA}}

\author{Yun Huang}
\affiliation{%
  \institution{School of Information Sciences, University of Illinois at Urbana-Champaign}
  \city{Champaign}
  \state{Illinois}
  \country{USA}}


%\affiliation{%
%  \institution{The Kumquat Consortium}
%  \city{New York}
%\email{jpkumquat@consortium.net}

\renewcommand{\shortauthors}{Liu et al.}

% \begin{abstract} %138 words
% Recent advances in multi-agent systems (MAS) enable users to have dialogue with AI-simulated personas to assist with complex knowledge work. However, most work centers on single-agent interaction, and how users perceive and control multiple personas remains underexplored.
% We present \systemName, a novel multi-agent system that allows users to control persona selection and visualize structured deliberation among LLM-simulated domain experts in the context of research ideation.
% Eighteen participants used \systemName to develop interdisciplinary research proposals, compared to a group-chat baseline with limited persona-control.
% Our findings show that \systemName's design elicited significantly more higher-order critical-thinking behaviors (e.g., inference, analysis, synthesis, and application) and  interdisciplinary discussions, followed by more  proposal revisions with improved clarity and detail. Meanwhile, the less-controlled group chat was preferred for interpreting well-defined knowledge. These  findings provide empirical evidence and design implications for balancing user control and agent orchestration.
% \end{abstract}

\begin{abstract} 
Recent advances in multi-agent systems (MAS) enable tools for information search and ideation by assigning personas to agents. However, how users can effectively control, steer, and critically evaluate collaboration among multiple domain-expert agents remains underexplored. 
We present \systemName, an interactive MAS that visualizes and structures deliberation among LLM agents via a forum-style interface, supporting @-mention to invite targeted agents, threading for parallel exploration, with a real-time mind map for visualizing arguments and rationales.
In a within-subjects study with 18 participants, we compared \systemName\ to a group-chat baseline as they developed research proposals.
Our findings show that \systemName\  significantly increased the frequency and depth of critical-thinking behaviors, elicited more interdisciplinary replies, and led to more frequent proposal revisions than the group chat condition. We discuss implications for designing multi-agent tools that scaffold critical thinking by supporting user control over multi-agent adversarial discourse. 
\end{abstract}


\begin{teaserfigure}
  \includegraphics[width=\textwidth]{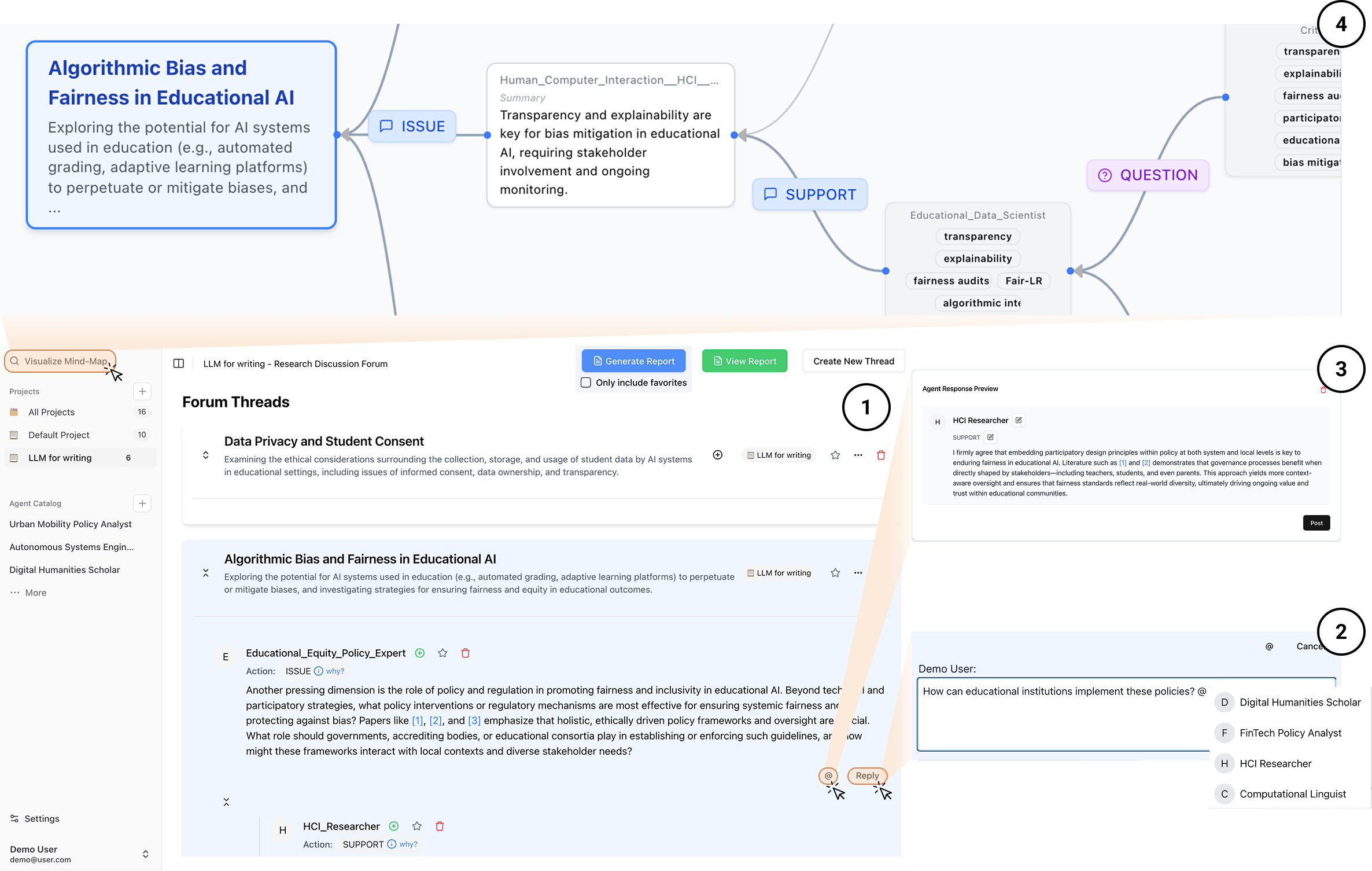}
  \Centering
  \caption{
  \systemName \:interface. The system supports the following features: \ding{172}  Upon login,  the main dashboard presents a forum-style interface that enables reply threading and structured multi-agent deliberation. \ding{173} Clicking the reply button allows users to respond to a post in the forum and tag additional agents into a single thread discussion. \ding{174} Clicking the button next to “Reply” activates the what-if feature, allowing users to explore hypothetical responses from agents. \ding{175} A mindmap feature provides a visualization of agents’ posts, replies, and deliberation actions with rationales, including semantic zooming to support sensemaking. %\yun{it would be great if the top map graph can be taller, including more nodes.} \yun{double check forum-style is used in the paper, without d.} 
  }
  \Description{
  Screenshot of Perspectra showing coordinated mind-map and forum views. Top: a mind map centered on “Algorithmic Bias and Fairness in Educational AI,” with ISSUE, SUPPORT, and QUESTION branches. A highlighted card summarizes “Transparency and explainability are key…”; an “Educational Data Scientist” node lists terms like transparency, explainability, fairness audits, and algorithmic interventions. Bottom: the forum interface. Left sidebar lists Projects and an Agent Catalog; the main pane lists threads, including “Data Privacy and Student Consent” and “Algorithmic Bias and Fairness in Educational AI.” In the latter thread, an Educational_Equity_Policy_Expert posts an ISSUE with numbered citations; an HCI_Researcher replies with SUPPORT. Callouts: (1) circles the Reply action on a post; (2) shows the user compose box asking “How can educational institutions implement these policies?” with selectable personas (Digital Humanities Scholar, FinTech Policy Analyst, HCI Researcher, Computational Linguist); (3) highlights an “Agent Response Preview” panel with a drafted HCI_Researcher message and Post button; (4) points to the mind-map visualization toggle.
  }
  \label{fig:teaser}
\end{teaserfigure}


\maketitle

% under a variety of application domains such as scientific research, healthcare, and education~\cite{Baek2024ResearchAgentIRA,chen2025mdteamgpt,zhang2024simulating}
 % a need that is increasingly crucial amid concerns about diminishing critical thinking skills with emerging LLM tool use~\cite{lee2025impact}.
\section{Introduction}
% \yiren{Highlight critical thinking more when motivating the research?}
LLM-based Multi-Agent Systems (MAS)~\cite{guo2024large} are increasingly being adopted across application domains because of their ability to perform more complex tasks than single-agent pipelines through collaboration among multiple agents ~\cite{Baek2024ResearchAgentIRA,chen2025mdteamgpt,zhang2024simulating}. Such dynamics in MAS have the potential to benefit both observers and overall task completion. This is similar to human collaboration, where dialogue among human experts promotes critical thinking in interdisciplinary learners~\cite{nussbaum2008collaborative,weber2023fostering,imbruce2024raising,lu2020effects,mcnair2019peer} and early-stage researchers~\cite{weber2023fostering,imbruce2024raising}. Despite this potential benefit, prior research has largely focused on designing effective communication with a single AI agent~\cite{huang2025deep,zheng2025deepresearcher} for complex knowledge work, leaving underexplored the opportunities and challenges of how users perceive and control multiple agents in collaborative settings. 
%More specifically, there exists a gap in understanding how MAS interaction designs can move beyond passive intake of multi-agent outputs (e.g., deep research reports \citep{zheng2025deepresearcher}) to actively engaging users in the collaborative process, thereby granting user with more control, agency, and opportunities to practice critical thinking skills.
% MOTIVATE CRITICAL THINKING
%Designs for MAS should not only focus on generating high-quality outputs, but also support user engagement and learning through collaboration with the AI agents.

% \pao{State the challenges of controling multiple agent here before the question? Framing: big benefits of multiagent (already stated), but it is also harder to control (if there is evidence of this being hard it raises up the problem)} \ta{I might move the sentence on challenges for designing interactive MAS up here instead and keep it broader than research ideation. Those seem like the key (broad) challenges of MAS}
Challenges in designing interactive MAS generally involve three issues: 1) the cognitive burden required for users to select and coordinate agents~\cite{Zuo2025LargeLM,wang2025large} (which agents to call); 2) information overload over parallel generation~\cite{Zhang2025GMemoryTH}; and 3) difficulty interpreting agent actions and rationales~\cite{Schmbs2025FromCT,zhang2025agent}. 
In this paper, we ask: \textit{how to better support users to control multi-agent collaborations?}
% \yun{need to justify why selecting interdisciplinary research ideation as a study context. something like} 
Specifically, we choose to explore this question in the context of how enhanced control can improve ideation quality and critical thinking when conducting interdisciplinary literature reviews. This is because interdisciplinary research fundamentally requires researchers to reconcile differing assumptions, methods, and disciplinary norms~\cite{liu2025personaflow}. Multi-agent systems with agents representing unique perspectives from different expert domains have been shown to be effective at enabling cross-field knowledge integration, increasing the richness of the interaction process, and stimulating critical thinking~\cite{liu2024personaflow,Zhang2025NovelSeekWAA}. However, these systems did not allow users to control multi-agent deliberation on different topics simultaneously.

% \ta{also what evidence do we have that these are two major gaps in interdisciplinary research? }\yiren{added mention of the two rounds of pilot studies}\pao{these don't sound like gaps (sth is missing). Could we phrase it more as "lack of ..."?}\yiren{revised} \ta{right now it looks like you are identifying two sets of gaps or issues, the first three and then the seciond two from the pilots, can you introduce one set that motivates the work here?}
To better understand the support users desire in such collaborations, we first conducted two rounds of pilot studies that uncovered key challenges perceived by users during ideation in interdisciplinary research contexts: (a) a lack of user control to steer the discourse among agents and probe emergent sub-topics, and (b) difficulty comprehending the evolving structure of multi-agent discussions, which increases cognitive load and impedes sensemaking.
%To investigate how different interaction designs can address these  challenges, we empirically compared two MAS interaction patterns that facilitate different levels of user control: 1) a forum-style, panel-like design, called \systemName, with targeted \textsf{@-mentions}, branching threads, and interactive visualization of deliberation actions, and 2) a group-chat, flat design with broadcast prompts and a single linear stream. 

We then designed and implemented \systemName, which provides two major design components to support user control and sensemaking of multi-agent deliberation: 1) \textsf{@-mention} and \textit{reply} that allow  users to participate in and invite user-chosen agents into an ad-hoc conversation, 
%to participate in discussion with persona-driven agents and perform persona selection as ways to steer the discussion by participating and inviting user-chosen agents into an ad-hoc conversation,
meanwhile revealing the reasoning processes of domain-expert agents during a group discussion through interactive visualizations; and 2) thread branching that facilitates parallel exploration of multiple topics while allowing users to easily blend and remix outputs from agents with different backgrounds. 
These designs are inspired by traditional forum interfaces, which are commonly used to gather feedback from individuals with interdisciplinary expertise and perspectives (e.g., ResearchGate~\cite{thelwall2017researchgate,zhang2024form,tian2021system} and Academia.edu~\cite{ovadia2014researchgate}). We extend this approach by allowing users to dynamically control which personas to engage with on the fly. 
In order to reduce cognitive load and improve sensemaking across threads, \systemName\ also provides visualization of agent deliberative actions and rationales through a dynamic mind map feature that assists navigation. 
To evaluate the effectiveness of \systemName\ design, we also implemented a group chat interface as the baseline condition. The group-chat design allows the user to chat with agents playing multiple personas, in one session, without being able to create different threads or (de-)select personas for sub-topics. This baseline approximates mainstream usage of systems such as ChatGPT, Claude, Grok, and Gemini, in which users can request persona shifts within a single chat session. %with limited user control over multi-agent collaboration that simulates existing design of multi-agent knowledge discovery and ideation systems~\cite{jiang2024into}.
% The forum-style design is informed by online discussion forum widely used by many to gather feedback for knowledge work requiring interdisciplinary expertise and perspectives (e.g., ResearchGate~\cite{thelwall2017researchgate,zhang2024form,tian2021system} and Academia.edu~\cite{ovadia2014researchgate}). Such communication modality also offers unique turn-taking mechanism~\cite{zhang2024form,tian2021system} that can be utilized in human-agent interaction, which are found to be beneficial for interdisciplinary exchange and ideation between human users. 
% \ta{much of this and the previous paragraph are repetitive}\yiren{removed most of them, and added forum motivation to the beginning of the paragraph}

% \yiren{More detailed explanation on what "agent perspective" actually means}

% We examined two system designs with varied levels of user control over multi-agent collaboration --- 1) the proposed \systemName\ system, and 2) the baseline group chat system with limited user control over multi-agent collaboration. 
We conducted a within-subjects study with 18 participants, who were asked to use both \systemName\ and the baseline group-chat design to develop a brief proposal on an interdisciplinary research topic. We analyzed system log, participants' survey scores, think-aloud data, and interview feedback. We compared the two designs in terms of proposal quality, required revision effort, observed interaction patterns, and the prevalence of critical-thinking activities.

Our findings make novel and timely contributions to the HCI community:
\begin{itemize}
    \item \textbf{\systemName, a new system design.} Support users to select personas and steer multi-agent deliberation. Features such as \textsf{@-mention} and thread branching allow users to create ad-hoc panels for tackling unfamiliar topics, while visualization of discussion structure, such as ISSUE, CLAIM, SUPPORT, REBUT, and QUESTION, assists their sensemaking.

    \item \textbf{Empirical evidence of enhanced critical thinking.}
   The experimental results show that \systemName\  significantly promotes critical-thinking activities, e.g., \textit{Application}, \textit{Analysis}, \textit{Inference}, and \textit{Evaluation}, compared to a group-chat baseline.  Participants initiated panel-like discussions to engage cross-disciplinary perspectives, examine assumptions, and refine interpretations, activities essential for higher-order reasoning.

   \item \textbf{Measurable improvements in proposal revisions and quality.} Participants using \systemName\ revised their proposals more frequently and achieved greater improvements in proposal quality compared to when using the group-chat condition. This finding suggests \systemName's practical impact, potentially going beyond interaction patterns to influence real work outcomes.

 \item \textbf{Discovery of designed and emergent affordances.} We observed both expected and emergent user behaviors. Designed affordances, such as \textsf{@-mention} driven deep dives with both a single agent and multi-agent sensemaking via the mind map, were used as intended. Beyond these design goals, participants created emergent practices such as leaving TODO-anchors and performing verification checks, showing the system’s flexibility, and new design opportunities for user-driven adoption.

    %enhanced control and diverse perspectives of the forum condition promotes critical thinking activities compared to the group chat interface, through user-initiated panel-like discussion using the \textsf{@-mention} feature to drive cross-disciplinary discussion; participants also made more proposal revisions and achieved better improvement on proposal quality when using \systemName\ to help revise their initial research proposal; We also observed common user affordances, such as using \textsf{@-mention} and reply to dive deep with a single agent and sensemaking interactions between multiple agents using the mindmap, and also emergent affordances outside the original design goals, including proactive engagement and verification through leaving TODO anchors during proposal edits; Moreover, we found that participants engage in complementary sets of critical thinking activities when using the two conditions --- users tend to engage more in activities such as questioning validity, examining assumptions, and refining interpretations when using \systemName, where the group chat condition were more often used for simpler information-seeking queries or requests that can be delegated to agents. 
    
    \item \textbf{Design implications for future systems.}  Our findings offer concrete design implications for knowledge-intensive ideation tools. We argue for enabling \textit{adversarial or dissenting agent responses} to foster critical reflection and for developing \textit{hybrid interaction models} that balance user control with agent autonomy, preventing cognitive overload while still supporting rich deliberation.
%Novel design implications for future ideation and knowledge-intensive deep search systems, arguing for enabling adversarial responses from agents during ideation (as opposed to sycophant responses) and a hybrid design to balance between user control and agent autonomy. 
\end{itemize}

\section{Related Work}

\subsection{Multi-Agent Systems for Research and Ideation Feedback Solicitation}
Many recent studies have considered the use of LLM-based agents to be an effective method for research ideation~\cite{Baek2024ResearchAgentIRA,Zhang2025NovelSeekWAA,Garikaparthi2025IRISIRA}.
Recent advancements in LLM-related research have explored how multi-agent systems can be applied to facilitate ideation in various domains and gather feedback~\cite{naik2025exploring}.  
The architectural foundation of multi-LLM ideation systems centers on role specialization and coordinated collaboration between distinct agents, which decomposes complex tasks into subtasks handled by specialized agents (e.g., ``Scientist,'' ``Critic'') to improve accuracy, completeness, and idea diversity~\cite{liu2024personaflow, Luoetal2025, Ghafarollahietal2024, Luetal2024, Uedaetal2025}. Coordination strategies in these systems range from user-orchestrated frameworks that prioritize user control~\cite{Parketal2023} to automated approaches where agents self-organize to solve problems~\cite{Ghafarollahietal2024}. To manage the complexity of these interactions and reduce cognitive load, recent work has focused on visual coordination tools and structured interfaces. These tools help users design and explore collaboration strategies visually~\cite{Panetal2024} and shift from reactive dialogues to more proactive, structured interactions with the multi-agent system~\cite{pengetal2024, wuetal2021, kimetal2023, maetal2023}.
There also has been work on using conversational agents to engage in community discussions~\cite{seering2019beyond}. 
\citet{Lietal2023CAMEL} proposed a multi-agent approach to simulate a society of LLM agents by allowing them to communicate with each other, where the dialogue and interactions can later be used for understanding agents' behavior and reasoning processes.

While prior work has focused on the architecture and coordination of multi-agent systems, less discussion has centered on how to offer fine-grained user control to selectively compose subsets of agents for emergent sub-topics, as interaction is often broadcast to all agents or fully automated in current MAS implementations~\cite{Parketal2023,Ghafarollahietal2024}.
We address these gaps with an interaction design that combines a forum-style interface with user control over agent selections through \textsf{@-mentions} for ad-hoc panel formation and a visualization of deliberative moves (e.g., claim, support, question) to surface stance relations and reasoning. Our design also reframes multi-agent output from a flat message stream into a navigable argument structure for scaffolding users' ideation processes across parallel topics.

\subsection{Balancing Learning and Cognitive Load: Collective Discourse and Distributed Cognition}
Recent studies leverage adversarial stances in conversational agents to provoke counter-arguments and promote critical thinking in groups. For design ideation, such agents help reduce design fixation by actively challenging dominant proposals~\cite{lee2024conversational}. In group decision-making contexts, devil's-advocate agents amplify minority voices to counter conformity pressure~\cite{lee2025amplifying} and mitigate social influence to improve deliberative quality~\cite{lee2025conversational}. Additionally, research has shown that structured deliberation among experts can be valuable for interdisciplinary learners~\cite{weber2023fostering,imbruce2024raising}. Similar discourse can also be found in research related to inquiry-based learning~\cite{lu2020effects,mcnair2019peer} and in the context of argumentation-centered collaborative peer learning~\cite{nussbaum2008collaborative}. 
A complementary perspective from cognitive neuroscience suggests \textit{Cognitive Synergy} as a driver of complex human cognition~\cite{luppi2022synergistic}. This suggests that systems should foster explicit combination of diverse and complementary perspectives.
Recent research has also heavily discussed applications of LLM agents in performing knowledge-extensive information retrieval tasks, with a well-established application example of deep research agents~\cite{huang2025deep,zheng2025deepresearcher}. This methodological paradigm emphasizes the use of LLM agents, often multiple and parallelized, to assist users in exploring and synthesizing large volumes of information into report. However, a challenge remains in how to 1) effectively grant users agency and control over the search process, and 2) how to present the information in a way that is easy for users to understand and evaluate.
Without a clear view of how agents reason, debate, and build upon each other's ideas, users may struggle to critically evaluate the generated feedback. 

Distributed Cognition theory~\cite{hutchins1995cognition} has been applied in a wide range of system designs to support sensemaking ~\cite{andrews2010space,paul2009cosense} through the breakdown of complex cognitive tasks into subtasks.
We build on these insights by applying guidance of explicit argument structures in multi-agent deliberation to scaffold users' critical evaluation during ideation, by providing visualization of agents' deliberation acts in a LLM-based ideation system to reduce cognitive load and nudge active reasoning instead of passive consumption~\cite{Panetal2024,pengetal2024}.

\subsection{Critical thinking activities in interdisciplinary learning}
\label{sec:relatedwork-criticalthinking}
Critical thinking skills and activities have been widely studied in educational psychology and pedagogy, particularly in the context of interdisciplinary learning~\cite{facione2011critical}.
We consider these categories as higher-order critical thinking activities because they require learners to move beyond recall and literal comprehension to generate, transfer, and judge knowledge. Bloom's taxonomy~\cite{bloom1956taxonomy} has been widely used to classify different types of cognitive activities, including \textit{knowledge}, \textit{comprehension}, \textit{application}, \textit{analysis}, \textit{synthesis}, and \textit{evaluation}.
In the cognitive domain, \textit{Application} entails using concepts and procedures in novel contexts, and \textit{Evaluation} requires making warranted judgments against criteria~\cite{bloom1956taxonomy,wilson2016anderson}. Contemporary critical-thinking accounts likewise identify \textit{inference} and \textit{evaluation} as core operations of expert judgment, which draws justified conclusions from evidence and assessing the credibility and quality of claims~\cite{facione1990critical}. 
Although the taxonomy lists ``inferring'' under understanding, the operation involves integrating prior knowledge with incomplete information to construct meaning not explicitly stated, which aligns with higher-order reasoning in practice. Empirical assessment work also groups tasks requiring \textit{application}/\textit{analysis}/\textit{evaluation} as higher-order because they demand integration and transfer to unfamiliar problems~\cite{jensen2020testing}. 
% critical thinking in learning context works + deliberation ...

Critical thinking is also essential for developing scientific literacy and fostering learning~\cite{facione2011critical}.
Recent research has revealed the risk of the use of GenAI technology in knowledge work reducing critical thinking skills~\cite{lee2025impact}. Research has begun investigating how to design GenAI systems that can support critical thinking activities~\cite{liu2024personaflow}.
In this work, we explore designs that nudge participants to engage in critical thinking activities. More specifically, we design agents' deliberation actions and the overall interaction model to promote active reasoning from users, targeting the forms of reasoning closely associated with critical thinking, rather than mere recall and comprehension.

% \yiren{may check out literature related to liberal law school education works on critical thinking + interdisciplinary; medici effect book}

\section{Methods}

\subsection{Iterative Design through Two Rounds of Pilot Studies}
\label{sec:iterative-design}
% Our initial design of \systemName\ was informed by past literature and theories. 
% The literature highlighted users' need to address the challenges of interdisciplinary research exploration, particularly in the context of collaborative discourse and critical thinking. 

% Past research has explored the use of LLM-based agents with different personas and profiles to support open-task knowledge discovery~\cite{jiang2024into}.
% However, these prior approaches face several challenges. Users often experience high cognitive load when reading extensive long-form text generated by multiple agents. Additionally, users struggle to track discussion context and understand the dynamics of multi-agent conversations as they unfold. There is also frequently a lack of transparency regarding the rationales behind agent interactions, making it difficult for users to understand why agents respond to each other in particular ways. These challenges limit the effectiveness of multi-agent systems for supporting complex interdisciplinary research exploration.

To address the gap of exploration across multiple perspectives in a deliberation setting, we introduce \systemName.
The design of \systemName\ is informed by an iterative design process that involved two rounds of prototyping and user feedback. 
We conducted two rounds of pilot studies with a total of 8 participants, including researchers from various disciplines. 
We first introduce an initial design of \systemName, drawing inspiration from online forum discussion layouts and interaction designs, which have been shown to facilitate collaborative ideation~\cite{tian2021system} by supporting context tracking and sensemaking in forum-based online discussions~\cite{zhang2024form}.

% \yiren{The connection to the DC part needs to be made clearer} 
During the first round of pilot, we presented a low-fidelity prototype to users and gathered their feedback and thoughts on the design. The prototype featured a forum-style interaction design where users could engage in threaded discussions with multiple expert agents by replying to a post or reply.
We found that users generally expressed interest in the online forum-based interaction. However, they identified several pain points, mostly related to information overload, difficulty in tracking discourse context, and challenges in accurately understanding agents' rationales behind responses.
Participants suggested several improvements, including: 1) a feature to allow users to more easily explore and verify the literature sources and rationales each agents referenced during discussions, for both purposes of transparency and sensemaking of agents' background (\userquote{You can also add the list of papers mentioned... For me, finding good relevant papers is really hard.} --- I1); 2) a mechanism to support easier navigation of discussion context and structure (\userquote{I kind of feel like the interaction between several people in the conversation might be a little bit... complex.} --- I3).
% \yiren{Add some quotes from users; also need rationales to support agent paper list and memory view}

\begin{figure}[t]
    \centering
    % Side-by-side iterative prototype evolution
    \begin{minipage}[b]{0.48\linewidth}
        \centering
        \includegraphics[width=\linewidth]{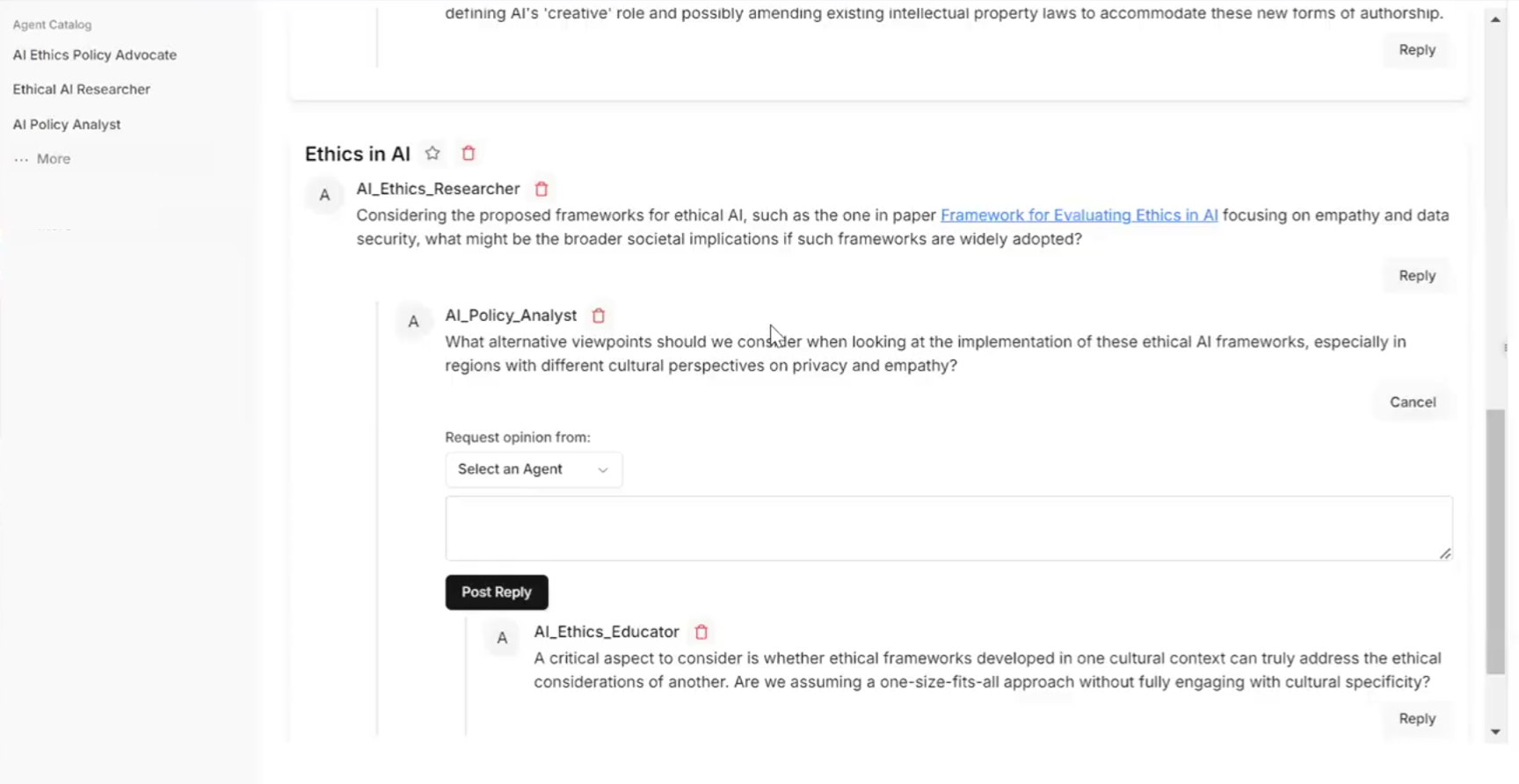}
        \vspace{2pt}
        \caption*{(a) Initial low-fidelity prototype (Pilot Round 1) with basic threaded forum interaction.}
        \Description{The Perspectra interface displays a forum thread titled ‘Ethics in AI.’ An AI_Ethics_Researcher post cites a ‘Framework for Evaluating Ethics in AI’ and asks about broader societal implications. A nested reply from AI_Policy_Analyst raises cultural differences in privacy and empathy. Below the thread is a compose box with a ‘Request opinion from — Select an Agent’ dropdown and a ‘Post Reply’ button; another reply from AI_Ethics_Educator cautions against one-size-fits-all ethical frameworks. A left sidebar labeled ‘Agent Catalog’ lists personas such as AI Ethics Policy Advocate, Ethical AI Researcher, and AI Policy Analyst, with a ‘More’ option.}
    \end{minipage}\hfill
    \begin{minipage}[b]{0.48\linewidth}
        \centering
        \includegraphics[width=\linewidth]{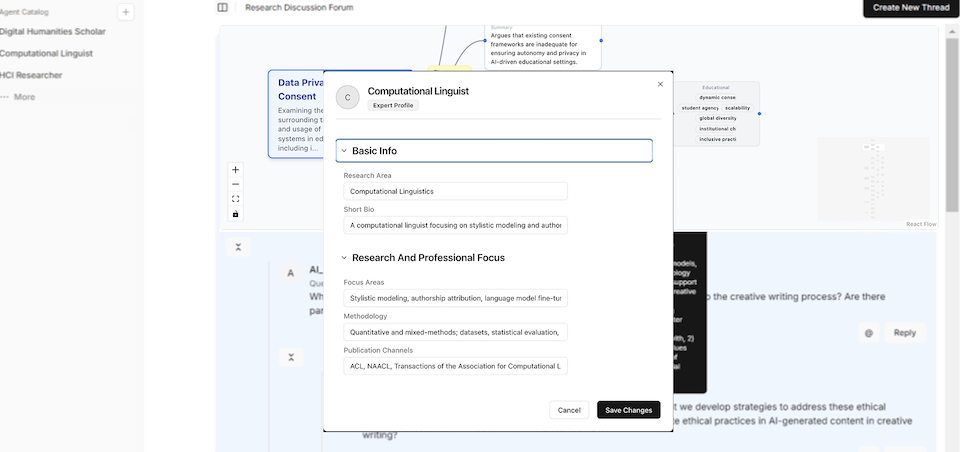}
        \vspace{2pt}
        \caption*{(b) Refined prototype (Pilot Round 2) with mind map and agent profiles}
    \end{minipage}
    \caption{Iterative prototype evolution of \systemName: (a) the initial forum-style design used to elicit early feedback; (b) the refined design incorporating participant suggestions to reduce information overload, surface agent rationale sources, and improve navigation of discussion context.}
    \Description{The interface shows a ‘Research Discussion Forum.’ Centered is a modal titled ‘Computational Linguist — Expert Profile’ with collapsible sections. ‘Basic Info’ lists Research Area: Computational Linguistics and a short bio. ‘Research and Professional Focus’ includes Focus Areas (e.g., stylistic modeling, authorship attribution, language-model fine-tuning), Methodology (quantitative and mixed-methods; datasets; statistical evaluation), and Publication Channels (e.g., ACL, NAACL, TACL). Buttons at the bottom read ‘Cancel’ and ‘Save Changes.’ Behind the modal, the left sidebar ‘Agent Catalog’ lists Digital Humanities Scholar, Computational Linguist, and HCI Researcher with a ‘More’ option. The background shows a mind-map node labeled ‘Data Privacy and Consent…’ with zoom controls, and part of a discussion thread below; the top-right has a ‘Create New Thread’ button.}
    \label{fig:iterative-design}
\end{figure}

Based on the feedback, we further refined the prototype and conducted another round of pilot study with a different group of participants (N=4). 
The second version of the prototype implemented improvements based on the feedback from participants during the first round, including 1) a mind map visualization to help users track the discussion context and structure; and 2) a panel that displays detailed agent profiles.
Our second round of pilot studies revealed several key areas for improvement. First, participants (I5, I6, I8) noted that the agents often lacked awareness of user intent during interactions, making it difficult to steer conversations in desired directions. Second, participants (specifically I5) expressed the need for clearer definitions and improved visibility of deliberation mechanics, as they sometimes struggled to understand why agents were making certain argumentative moves. Finally, multiple participants (I7, I8) requested mechanisms to more easily follow up with individual agents during parallel and multi-threaded discussions, suggesting that more direct engagement options with specific personas would enhance the deliberation experience. These insights guided our final system refinements before the formal user study.
The two rounds of pilot studies informed two major design goals for the system:
\begin{itemize}
    \item \textbf{DG1 (Choosing Expert Personas to Steer Forum Discourses)}: Enable users to dynamically involve and steer the discourse between multiple agents to deepen dialogue around emerging topics.
    \item \textbf{DG2 (Structuring Dialogue for Effective Comprehension)}: When there are multiple agents involved in discuss, the interface should facilitate users' sensemaking of evolving discussion and its dynamics by reducing cognitive load, e.g., through formalization and visualization of discourse actions.
\end{itemize}

% \yun{including: 1) formalizing dialogues between agents. 2) Used models are part of computational argumentation theory, aiming to let agents argue using well-defined rules.}
% \item \textbf{DG1}: The system should support Users to easily understand discussion structure and dynamics (section 3.3 xxx, yy, zzz) and make sense of the discourse. figure 2 (action type), mind map

\subsection{System Design}
\label{sec:system-design}

We propose \systemName, an interactive system designed to enhance critical thinking through structured multi-agent deliberation and visualization of discussion dynamics. In the section, we describe the design of \systemName's key components that address the design goals outlined in~\cref{sec:iterative-design}.

\subsubsection{Agent Interaction and Perspective Exploration (\textbf{DG1})}
\systemName\ allows users to interact with multiple expert agents through a threaded interface as if in a panel discussion setting involving multiple researchers with their own domain background knowledge. Each agent is assigned a specific background profile, and users can engage in interaction with agents through several mechanisms designed to facilitate discourse, including:

\begin{figure}[t]
    \centering
    \includegraphics[width=\linewidth]{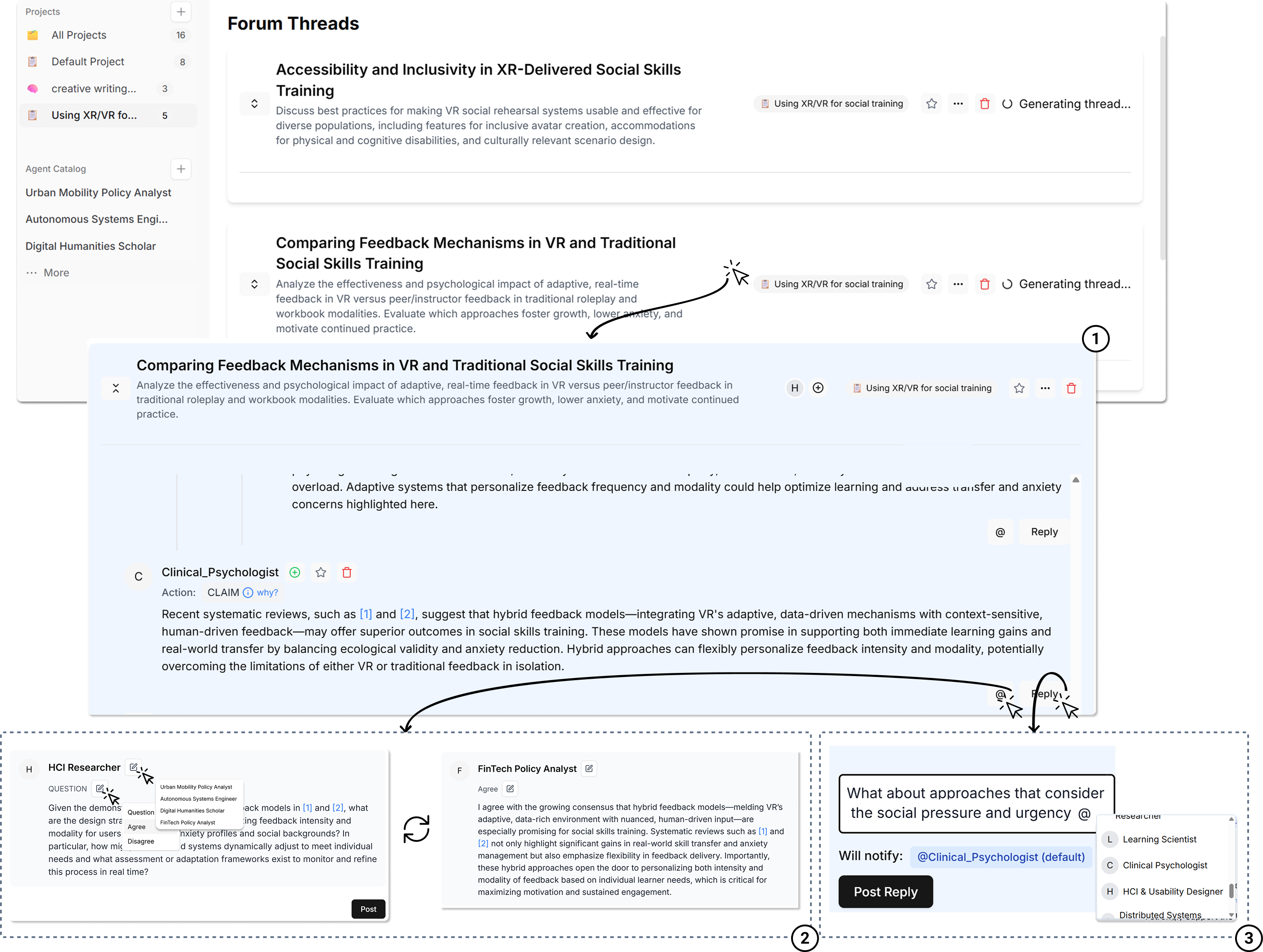}
    \caption{\systemName~system interface utilizes forum-style layout to support unique turn-taking deliberation interactions between agents, and visualization of threaded parallel topics. The interface supports interactions including: \ding{172} clicking on a discussion thread to collapse/expand the thread to hide/show detailed replies; \ding{173} clicking on the @ button under each reply allows generation of a ``what-if'' panel showing hypothetical replies from a chosen agent and stance combination, the user can also change the agent or stance with the panel to re-generate the reply; \ding{174} clicking on the reply button spawns a text input box in which the user can type their reply, and the user can also type @ to select from a list of available agents to be included in follow-up discussion.
    }
    \Description{An illustration of a collaborative online platform with a clean, modern interface. The main section displays a forum thread titled "Comparing Feedback Mechanisms in VR and Traditional Social Skills Training." The thread shows comments from different users, including "Clinical_Psychologist" who is proposing a new approach. The bottom of the image shows three smaller, overlapping pop-up windows: one with a user named "HCI Researcher" asking a question, a second with "FinTech Policy Analyst" agreeing with a previous comment, and a third showing a blank reply box with the text "What about approaches that consider the social pressure and urgency?" Callout 1 highlights the open thread view; Callout 2 points to the ‘Reply’ action; Callout 3 shows a compact reply composer with @mention and ‘Will notify: @Clinical_Psychologist (default)’ above a ‘Post Reply’ button.}
    \label{fig:system-interface-forum-layout}
\end{figure}

\begin{enumerate}
    \item \textbf{Free-text replies}: As shown in~\cref{fig:system-interface-forum-layout}, users can reply to any agent's post, with @-mentions to involve specific agents in the conversation. Once a user's reply is posted, all mentioned agents will respond to the user's reply. If no additional agents are tagged, the agent that made the target post will respond.
    \item \textbf{Action requests}: The system also provides a quick action for users to select a particular agent and take a specific stance when generating the response (i.e., agree, disagree, question). We chose to adopt a simplified list of stances instead of the full deliberation actions (as in \cref{sec:deliberation-framework}) to reduce cognitive load for users.
    We chose to adopt a simplified list of stances instead of the full deliberation actions (as in \cref{sec:deliberation-framework}) to reduce cognitive load for users.
    Once the user clicks the action, the system displays a ``What-if'' perspective panel to help users explore alternative viewpoints on the same topic. 
    \item \textbf{Thread branching}: Users can also choose to create new discussion threads based on specific responses for deeper exploration of topics that emerge during the discussion, without being limited to the pre-generated topic, like in a typical chat interface.
\end{enumerate}

 %, directly supporting our goal of highlighting collaborative discourse with varied viewpoints (\textbf{DG1}).

% \subsubsection{Perspective Summary}
% \begin{figure}
%     \centering
%     \includegraphics[width=0.7\linewidth]{imgs/perspective_viewer_interface.png}
%     \caption{Interface of the perspective viewer.}
%     \label{fig:perspective-viewer-interface}
% \end{figure}

% \yiren{This is not used as much during the user study, maybe remove?}
% To help users synthesize insights from complex deliberations, \systemName\ provides dynamic perspective summaries that aggregate insights from existing threads or those the user has marked as favorites. This feature helps users identify common themes, different perspectives, and relationships between different disciplinary viewpoints, further supporting \textbf{DG1} by making diverse perspectives more accessible and comparable.

%\subsubsection{Persona Profile and Memory}
This interaction design facilitates users' exposure to diverse perspectives from different disciplinary backgrounds. \systemName\ further provides features to enable transparency and customization of the agent personas. For example, users can access and edit detailed profiles for each persona, which are initially derived from the Persona Hub dataset \cite{ge2024scaling}. This feature allows users to tailor agent backgrounds to better match their own research contexts or explore specific types of disciplinary perspectives. More details about the implementation of agents and personas can be found in \cref{sec:agent-backend}.
Each agent simulates a researcher with a particular domain background, and the agents have access to their own collections of literature and unique ways of thinking. 
More specifically, each agent persona is presented by:
\begin{enumerate}
    \item \textbf{Profile information} detailing expertise, background, familiar methodology of research, etc., accessible by clicking on an agent's name on the left panel on the interface.
    \item \textbf{Literature collection} accessible by clicking on an agent's avatar, revealing the sources of knowledge informing the agent's contributions.
    \item \textbf{Agent memory} through an alternative tab within the same panel as the literature collection that displays the agent's internal memory state, enhancing transparency about how the agent forms and maintains its perspective. The memory viewer provides two forms of visualization: 1) a chronological stream, and 2) a collapsible lineage tree to support user sensemaking of why an agent now argues a certain way. This design aims to increase transparency (users can inspect ``why this shift occurred''), which supports \textbf{DG1} by externalizing intermediate reasoning states, while avoiding overwhelming users with sensemaking of agent intentions through responses.
\end{enumerate}
Screenshots of the agent profile editor and memory viewer are provided in Appendix~\ref{fig:agent-interfaces}.
These transparency features are designed to help users understand why agents offer particular perspectives and how their backgrounds led to their responses, better supporting users' sensemaking of agents' messages.

\subsubsection{Threaded Forum and Mind Map for Multi-Expert Deliberation (\textbf{DG2})} 
The core component of \systemName\ is an interface that facilitates deliberation between LLM-based agents through a design that emulates an interactive online discussion forum. This design allows users to engage in multi-agent discussions, where each agent represents a different disciplinary perspective, as shown in~\cref{fig:system-interface-forum-layout}.
The system is structured around a threaded forum format, which organizes discussions around different topics into threads. Each thread is dedicated to a specific sub-topic, allowing users to focus on one aspect of the discussion at a time while providing the flexibility to switch between parallel threads.  The system also provides a project tab that allows users to create and manage threads around different individual research projects or ideas. Each project page can host multiple discussion threads. 
The system generates initial suggestions of thread titles and descriptions based on the user's initial research proposal input, which can be edited and confirmed by the user before starting the forum discussion. 
Grounded in Distributed Cognition theory~\cite{hutchins1995cognition}, our threaded forum design aims to assist users in decomposing complex topics into focused discussions to reduce cognitive load and enhance ideation, as supported by prior work~\cite{sweller2010cognitive,ma2025dbox,kim2022mixplorer}. This structure aims to help users explore specific sub-topics in depth while maintaining awareness of the broader discussion context.

%\subsubsection{Mind Map Visualization}

\begin{figure}
    \centering
    \includegraphics[width=\linewidth]{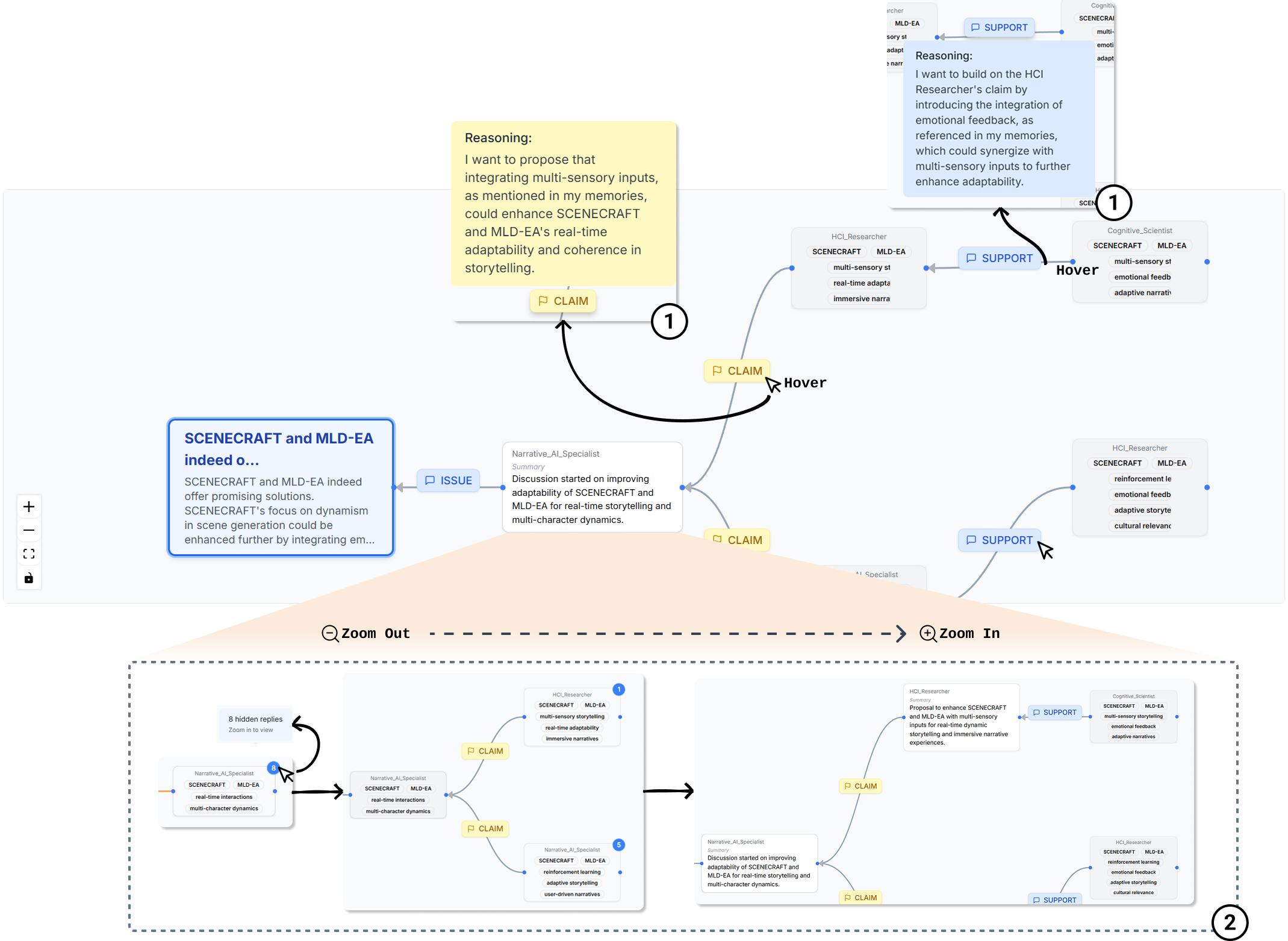}
    \caption{An interface of the mind map feature. Nodes represent posts/threads/replies with different levels of detail with semantic zooming, and edges encode agents' deliberation acts (details can be found in \ref{sec:deliberation-framework}). \ding{172} Hovering an action chip or its connecting edge surfaces an inline rationale card with the posting agent's reasoning. \ding{173} Semantic zoom shifts between the keyword view and the summary view for each node.}
    \Description{A user interface for a visual argumentation or concept mapping tool. The main screen displays a mind map with a central topic, "Narrative AI Specialist Summary." Various nodes labeled "CLAIM," "SUPPORT," and "ISSUE" branch from the center. A pop-up yellow sticky note labeled "Reasoning" is shown when hovering over a "CLAIM" node, explaining the user's thought process. A bottom panel illustrates the "Zoom Out" and "Zoom In" functionality by showing the map at different levels of detail.}
    \label{fig:mindmap-interface}
\end{figure}

In order to offer a more organized and coherent visualization for deliberation dynamics,  \systemName\ also includes an interactive mind map visualization that represents the discussion in a graph-based layout, as shown in \cref{fig:mindmap-interface}. The nodes display thread/post-level information, while edges convey actions and rationale from each participating agent. This visualization provides an alternative, spatial representation of the deliberation structure.

The mind map implements semantic zooming capabilities as used by previous sensemaking support research~\cite{suh2023sensecape}, allowing users to adjust the level of detail displayed as they zoom in or out of the visualization. At a high level, users can see main argument structures and key points; zooming in reveals more detailed summaries of the original posts/replies. This feature is designed to directly support different levels of ideation and navigation needs. 
% \pao{One might wonder, why didn't we just use mind-map instead of the forum since it has all information + semantic zoom. One response could be, it is clunky to navigate and is good as a supplement view }
While the mind map offers a good overview of the deliberation's structure, we designed it as a supplementary view to the main forum interface. This is because we discovered during the second pilot study that a dense mind map can become hard to navigate for reading and authoring detailed responses compared to a nested, collapsible, threaded interface.
The mind map also incorporates context tracking, helping users to track back to the original content (post/thread) whenever a node is clicked on.
The mind map visualization is designed to support users in more effectively processing complex information across disciplines and topics, while still maintaining awareness of interconnections through the system's visualization components.

\subsection{Backend Implementation}

\subsubsection{Multi-Agent System for Research Ideation}
\label{sec:agent-backend}
We implement a multi-agent system that enables users to interact with multiple LLM-driven research personas implemented using AutoGen~\cite{wu2024autogen} framework. Each agent operates over shared tools and a graph-based retrieval database to support cross-perspective ideation while maintaining independent goals and memory.

\textbf{Agent components.} Each agent consists of three major components that guide its behavior: 
1) \textbf{Persona profile}: Persona profiles are populated from a structured taxonomy derived from \citet{ge2024scaling}'s work (e.g., discipline, methodological stance, research role, epistemic orientation, focus areas, methodology, publication channels, skills, and communication style) using agglomerative clustering;
The full taxonomy can be found in ~\cref{apdx:persona-taxonomy}. Persona prompts also include a brief background narrative to encourage coherent and diverse reasoning from each agent's own perspective;
% \yiren{Add more details about how the taxonomy was derived (mining from \cite{ge2024scaling}) and add evaluation, but too long probably put into appendix}; 
2) \textbf{Agent memory}: Beyond short-term memory (i.e., conversational history maintained by AutoGen), agents also have long-term cross-turn memory that uses a persistent store (implemented with a modified version of LangMem~\footnote{https://github.com/langchain-ai/langmem}) store;
Each agent periodically distills interaction context (its own prior posts, user prompts, retrieved literature summaries, and other agents' challenges) into compact research idea snippets. Each snippet captures a single evolving hypothesis, question, rationale shift, or methodological consideration, which can also reference earlier snippets it refines or extends. The memories are later referenced during each time of inference;
3) \textbf{Literature database}: A GraphRAG database (implemented using a modified version of LightRAG~\cite{guo2024lightrag} to support citation tracing) constructs and queries a knowledge graph over both entities and text snippets extracted from papers. The system also implements a paper search tool that queries the Semantic Scholar (S2) API~\cite{kinney2023semantic} and OpenAlex~\cite{priem2022openalex}. Retrieved papers are incrementally inserted into the GraphRAG database.

% \textbf{Discussion management.}
% A hierarchical DiscussionThreadManager records message trees, tool-call summaries, and multi-level summaries for efficient context reuse.
% \yiren{Elaborate}

\textbf{Agent reasoning and tool use.} 
Agents follow a ReAct-style reasoning loop~\cite{yao2023react} integrated with Autogen's tool-calling capabilities. In each turn, an agent first generates a \textit{plan} based on the current dialogue context and its persona, deciding whether to respond directly or use a tool. 
If a tool is needed, it executes an \textit{action}, such as querying the GraphRAG database for existing knowledge, searching external literature via the Semantic Scholar and OpenAlex APIs, or adding a newly found paper to its knowledge base. 
The raw output from the tool is then summarized and used to augment the agent's context. The agent then \textit{reflects} on this new information to revise its initial plan if necessary. For instance, if a literature search fails to support a planned argument, the agent might pivot to a different claim and conduct another round of search. Finally, the agent generates its \textit{response} based on the augmented context. To maintain a clean and interpretable deliberation history for the user, only the final agent rationale and a summary of the tool call are persisted, rather than the entire chain-of-thought process.

\subsubsection{Designing Agent Deliberation Framework}
\label{sec:deliberation-framework}
% \yiren{we design the system to encourage high-order thinking activities ...; then we tabulate the results and find some of the categories are more than others ... some are not ...}
We design the agents' action space by adopting several existing argumentation frameworks. Specifically, \citet{prakken2006formal} on argumentation speech acts, Toulmin's model of argument structure \cite{toulmin2003uses}, Walton \& Krabbe's dialogue types and commitment rules \cite{walton1995commitment}, and the Agent Dialogue Framework (ADF) \cite{mcburney2002games}. The goal of the design is to maintain expressive power for multi-agent deliberation while keeping the interaction readable to human readers' interpretation of agents' rationales.
We adopt a deliberation framework as agents' action space, following the typology of \citet{walton1995commitment} and the composition principles in ADF \cite{mcburney2002games}. Individual turns are realized using a small set of \textit{locutions} adapted from \citet{prakken2006formal} (e.g., assert, challenge, concede, retract) and \citet{toulmin2003uses} (Claim, supporting Grounds/Backing, potential Rebuttals). 
In the UI, these backend \textit{locutions} are displayed as labels (e.g., \textsc{Claim}, \textsc{Support}), the root threads always start with \textsc{Issue} posts (which open sub-threads/replies), and the mind map panel displays nodes (posts/replies) with edges labeled according to the deliberation acts (i.e., \textit{locutions}) that leads to each node, as a visualization of the structure of the deliberation. The action set and theoretical origin are shown as in~\cref{tab:action-set}.

\begin{table}[t]
\small
\renewcommand{\arraystretch}{1.1}
\setlength{\tabcolsep}{6pt}
\begin{tabularx}{\linewidth}{l X X}
\toprule
\textbf{Action} & \textbf{Definition} & \textbf{Theoretical Adaptation} \\
\midrule
\textsc{ISSUE} & Introduce a new question, sub-topic, or decision point. & Agenda setting in deliberation dialogue types \citep{walton1995commitment}; sub-dialogue launch schemas in ADF \citep{mcburney2002games} \\
\textsc{CLAIM} & State a position that the speaker commits to defend. & Assert, claim locutions and commitment updates \citep{prakken2006formal,walton1995commitment};  \textit{Claim} from~\citet{toulmin2003uses} \\
\textsc{SUPPORT} & Provide explicit support with argumentative content. & Toulmin \textit{Grounds}, \textit{Backing}, and \textit{Warrant} \citep{toulmin2003uses}; supplying reasons to ``why?'' challenges \citep{prakken2006formal} \\
\textsc{REBUT} & Provide a counter-argument that attacks a prior claim or support. & Toulmin \textit{Rebuttal} \citep{toulmin2003uses}; attack and defeat moves \citep{prakken2006formal} Premise/inference attacks collapsed for legibility \\
\textsc{QUESTION} & Ask for justification or clarification (``Why?'') about a claim. & Challenge, “why” locutions and burden transfer \citep{prakken2006formal,walton1995commitment} \\
% \textsc{CONCEDE} & Explicitly accept another agent's point. & Accept / concede moves updating commitment stores \citep{prakken2006formal,walton1995commitment}. \\
% \textsc{WITHDRAW} & Retract one of the speaker's own earlier points. & Retract / withdraw moves removing items from commitment store \citep{prakken2006formal,walton1995commitment}. \\
\bottomrule
\end{tabularx}
\caption{The design of action space for the agent deliberation driven by existing argumentation frameworks.}
\Description{The table defines five actions in agent deliberation and maps each to argumentation frameworks: ISSUE — introduce a new question, sub-topic, or decision point; tied to agenda setting in deliberation dialogues and sub-dialogue launch schemas in ADF (Walton 1995; McBurney \& Parsons 2002). CLAIM — state a position the speaker commits to defend; linked to assert/claim locutions and commitment updates, and Toulmin’s Claim (Prakken \& Sartor 2006; Walton 1995; Toulmin 2003). SUPPORT — provide explicit argumentative support; corresponds to Toulmin Grounds, Backing, and Warrant, and giving reasons to “why?” challenges (Toulmin 2003; Prakken \& Sartor 2006). REBUT — counter-argument attacking a prior claim or support; aligns with Toulmin Rebuttal and attack/defeat moves; premise/inference attacks are collapsed for legibility (Toulmin 2003; Prakken \& Sartor 2006). QUESTION — request justification or clarification (“why?”); mapped to challenge/‘why’ locutions and burden transfer (Prakken \& Sartor 2006; Walton 1995).}
\label{tab:action-set}
\end{table}

% \yiren{Need to double-think on the action set and the theoretical adaptation; Unify it}

% Why this framework for research ideation and knowledge discovery?
% Deliberation dialogues provide a natural scaffolding for exploring alternatives and trade-offs, while Toulmin-structured content and commitment-aware locutions make \emph{why} an idea is favored (or abandoned) transparent to readers. Combined with ADF’s support for embedding persuasion sub-dialogues within deliberation, the action set above yields discussions that are both procedurally well-formed (backend) and easily navigable (frontend), which we found crucial for helping users follow, audit, and build upon multi-agent analyses.

% \yiren{To finish}
% \yiren{Link each action to each original framework, } 
% \yun{given they are in the backend implementations, how are they linked to the interaction design?}

\section{User Study}
We designed a within-subject user study to evaluate the effectiveness and user experience using the \systemName. The user study focuses on validating the effectivess of \systemName\ and understanding how users engage with \systemName\ during ideation.
% More specifically, we aim to answer the following research questions:
% \begin{itemize}
%     \item \textbf{RQ1}: How does interacting with threaded, multi-agent interdisciplinary discussions influence participants' ability to identify and integrate diverse perspectives? 

%     \item \textbf{RQ2}: How do users use \systemName\ to develop research ideas?
    
%     \item \textbf{RQ3}: How do the features of \systemName\ foster critical reflection and evaluation during interdisciplinary research exploration?
% \end{itemize}

\subsection{Baseline} 
Past research has explored the use of multiple LLM-based agent for supporting information discovery by integrating insights from multiple perspectives and domains~\cite{jiang2024into}. 
We implemented a baseline condition that enables similar interactions to mainstream chat-based systems such as ChatGPT, Claude, Grok, and Gemini in order to evaluating and understanding our proposed interaction designs. 
We chose to implement the baseline instead of using existing systems (e.g., \cite{jiang2024into}) in order to separate the confounds from the additional interface and interaction design differences. 
The baseline offers a conversational interface with a single-textbox input. The interface allows users to chat with multiple agents in a vanilla group-based setting.
The agents are designed to respond to user queries in a similar manner as the agents used in \systemName\ with access to the literature search tool and GraphRAG database. However, the agents do not implement the deliberation action space as in \systemName in order to simulate the common implementations of MAS supporting information search.

\subsection{Study Procedures}
We conducted a within-subjects study with 18 participants to compare \systemName\ with a baseline multi-agent group-chat interface. The study was designed to evaluate how the proposed different interaction designs affected participants' critical thinking, perceived domain clarity, and research ideation outcomes.

\subsubsection{Participants and Recruitment.} We recruited 18 participants spanning undergraduate, master's, Ph.D., and post-doctoral researchers with interdisciplinary research experience. Participants have diverse domain background, and were recruited through university mailing lists and social media platforms. The study was conducted online through Zoom, and participants were compensated with \$20 per hour for their time. This study was approved by the university's Institutional Review Board (IRB). Details of participants' demographics are shown in~\cref{tab:participant-demographics}.

%TC:ignore
\begin{table*}[h!]
\centering
\footnotesize
\setlength{\tabcolsep}{6pt}
\renewcommand{\arraystretch}{1.15}
\caption{Participant demographics and research experience.}
\label{tab:participants}
\begin{tabularx}{\textwidth}{@{} l X l l @{}}
	\toprule
	\textbf{PID} & \textbf{Background} & \textbf{Education Level} & \textbf{Research Experience} \\
\midrule
T1  & Computer Science and Artificial Intelligence; Education and Learning Sciences; Human Computer Interaction & PhD Student         & 3--4 years \\
T2  & Computer Science and Artificial Intelligence; Education and Learning Sciences; Human Computer Interaction & PhD Student         & 3--4 years \\
T3  & Physics and Astronomy; Engineering and Technology; Computer Science and Artificial Intelligence; Social Sciences (e.g., Sociology, Anthropology); Psychology and Cognitive Science; Mathematics and Statistics; Humanities (e.g., History, Philosophy, Literature); Data Science and Information Technology & Undergraduate Student & 1--2 years \\
T4  & Physics and Astronomy; Computer Science and Artificial Intelligence; Mathematics and Statistics & Undergraduate Student & 3--4 years \\
T5  & Engineering and Technology; Psychology and Cognitive Science; Neuroscience and Behavioral Sciences; Data Science and Information Technology & Master's Student     & 1--2 years \\
T6  & Computer Science and Artificial Intelligence; Education and Learning Sciences & PhD Student         & 3--4 years \\
T7  & Engineering and Technology; Computer Science and Artificial Intelligence; Environmental Science and Sustainability & Master's Student     & 3--4 years \\
T8  & Biology and Life Sciences; Chemistry and Materials Science & Postdoctoral Researcher & 5+ years  \\
T9  & Computer Science and Artificial Intelligence; Law, Political Science, and Public Policy; Data Science and Information Technology & PhD Student         & 5+ years   \\
T10 & Engineering and Technology; Environmental Science and Sustainability & Master's Student     & 1--2 years \\
T11 & Biology and Life Sciences; Medical and Health Sciences; Social Sciences (e.g., Sociology, Anthropology); Law, Political Science, and Public Policy & Undergraduate Student & 3--4 years \\
T12 & Agricultural, Food, and Nutritional Sciences & PhD Student         & 3--4 years \\
T13 & Social Sciences (e.g., Sociology, Anthropology); Psychology and Cognitive Science; Economics, Business, and Management; Neuroscience and Behavioral Sciences & PhD Student         & 3--4 years \\
T14 & Computer Science and Artificial Intelligence; Education and Learning Sciences; Arts, Design, and Creative Studies & PhD Student         & 1--2 years \\
T15 & Computer Science and Artificial Intelligence; Education and Learning Sciences; Human Computer Interaction & PhD Student         & 3--4 years \\
T16 & Computer Science and Artificial Intelligence; Education and Learning Sciences & PhD Student         & 3--4 years \\
T17 & Computer Science and Artificial Intelligence; Human Computer Interaction & Undergraduate Student & 1--2 years \\
T18 & Medical and Health Sciences; Computer Science and Artificial Intelligence; Arts, Design, and Creative Studies; Data Science and Information Technology & PhD Student & 3--4 years \\
\bottomrule
\end{tabularx}
\Description{Table summarizing 18 participants' backgrounds, education, and research experience. Education: 10 PhD students, 4 undergraduates, 3 master's students, and 1 postdoc. Research experience: 5 with 1–2 years, 11 with 3–4 years, and 2 with 5+ years. Most common backgrounds include Computer Science & AI (12), Education & Learning Sciences (7), and Human–Computer Interaction (4), with additional fields across physics, biology/health, engineering, data science, environmental science, law/policy, social sciences, psychology, neuroscience, economics/business, and arts/design.}
\label{tab:participant-demographics}
\end{table*}
%TC:endignore

\subsubsection{Study Design and Protocol.}
We employed a counterbalanced within-subjects design where each participant completed two research ideation tasks, one using our \systemName\ system and one using the baseline group-chat interface. To control for learning and ordering effects, we counterbalanced the order of system presentation across participants. %We chose to develop a chat-only baseline interface within the same system framework (instead of using alternative systems~\cite{jiang2024into}) to isolate the effects of our proposed interaction designs from other confounding factors. 
\begin{figure}
    \centering
    \includegraphics[width=0.95\linewidth]{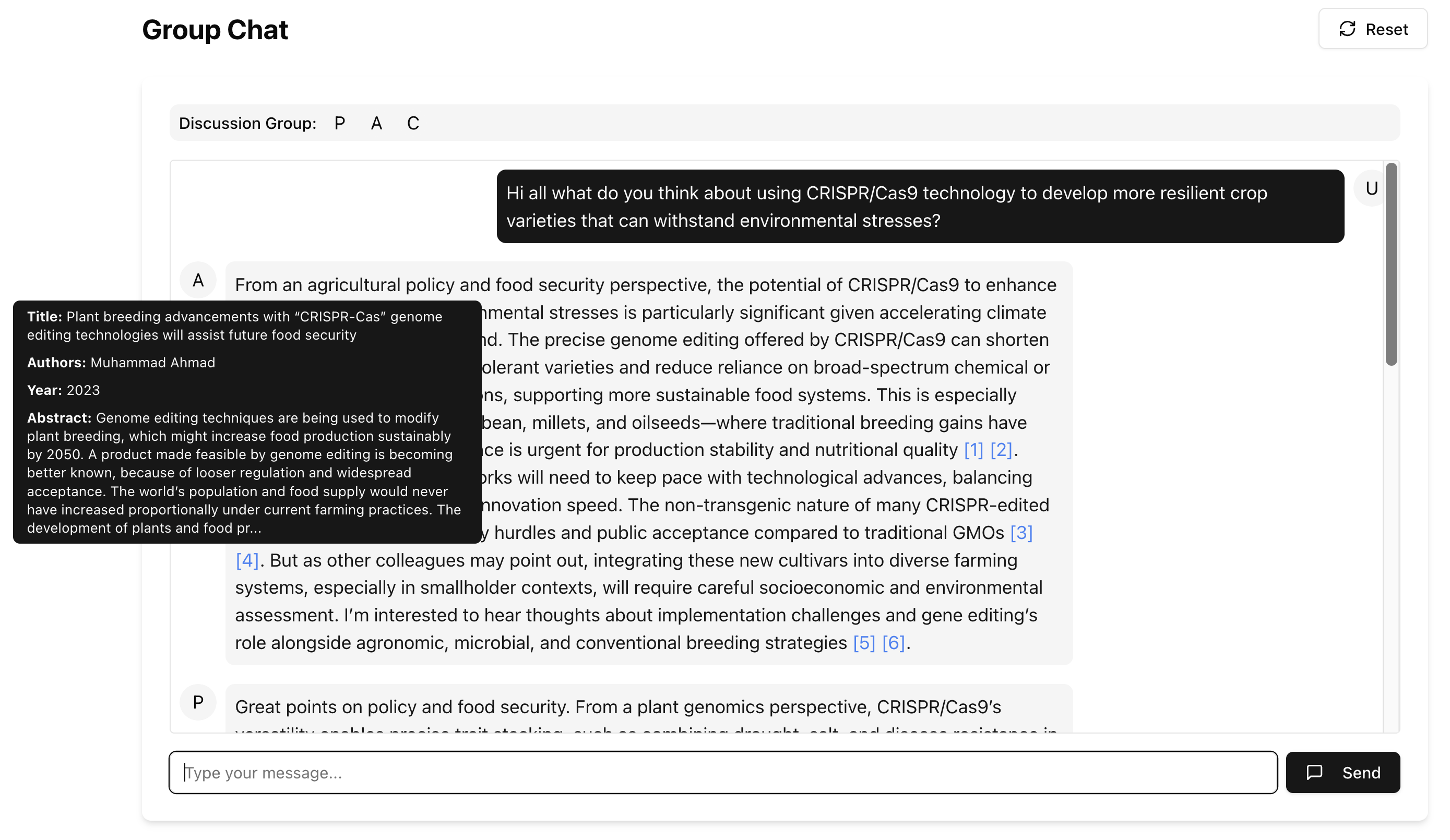}
    \caption{The interface of the baseline group-chat system used for the study. The baseline system implements a single-stream chatting interface without parallel-threading or persona (de)selection support, which approximates mainstream chatting applications. The system does have support for embedding and displaying citations.}
    \Description{Screenshot of a “Group Chat” screen. Below the header “Discussion Group: P  A  C”, the user (U) posts: “Hi all what do you think about using CRISPR/Cas9 technology to develop more resilient crop varieties that can withstand environmental stresses?” Participant A responds with a multi-sentence message about agricultural policy and food security, including inline numeric citations [1]–[6]. On the left, a black tooltip is open showing citation details for one source: title “Plant breeding advancements with ‘CRISPR–Cas’ genome editing technologies will assist future food security,” author Muhammad Ahmad, year 2023, and a truncated abstract. A message from participant P (plant genomics perspective) begins below A’s reply. At the bottom is a text box labeled “Type your message...” with a Send button; a Reset button sits in the upper-right corner. A vertical scrollbar is visible on the right.}
    \label{fig:baseline_interface}
\end{figure}

Each participant's study session lasted approximately 90 minutes.
Each session began with a 10-minute introduction where we explained the study goals and completed consent procedures. During this initial phase, we conducted a brief interview to gather information about participants' research backgrounds.
Before interacting with each system, participants completed a pre-session survey (7-point Likert scale) to establish baseline measures, including familiarity with the chosen topic, trust in GenAI, and self-assessed initial proposal quality and domain clarity.
The core part of the study consisted of two 30-minute system interaction sessions. In one session, participants used the baseline system featuring the group-chat interface for interacting with multiple AI agents. In the other session, they used \systemName\ with its threaded forum interface, mind map visualization, and other specialized components. Throughout both sessions, participants were instructed to think aloud, verbalizing their thoughts, reactions, and decision-making processes. These think-aloud sessions were recorded for later analysis.

For each system, participants followed a structured task sequence: they first entered their research idea in sections, including motivation, description of past research, methodology, and hypothetical findings, then engaged with the AI agents to explore their research topic. They used the respective system features to navigate different perspectives and disciplines, collecting insights to develop their research proposal. 
We use the written proposal as a proxy of ideation outcome in research ideation, as proposal writing entails the use of complex reasoning activities (e.g., problem framing, articulation of claims and rationales, synthesis of prior work, and methodological justification) to create an artifact aligned with established critical-thinking constructs~\cite{murtonen2019redefining,nussbaum2008collaborative,lu2020effects}. 
Before asking participants to engage with the systems, we first provided a detailed tutorial of each system's features and functionalities in the form of a walk-through using a sample research idea. During the study, users were encouraged to explore the system freely, but were also instructed to use each feature of the system at least once and ensure they engaged with key features of each system.
% \yun{be more specific on the tasks that they must do.}
After each system interaction, participants completed post-session surveys measuring cognitive load, usability, and critical thinking self-assessment (7-point Likert scale). The survey also collects the same set of measurements as in the pre-session survey, including self-perceived interdisciplinary clarity and proposal quality. Detailed survey items for both pre- and post-session surveys are provided in Appendix~\ref{apdx:survey-items}.
Each session concludes with a 15-minute semi-structured exit interview exploring participants' experiences with both systems. The interview focused on identifying ``aha moments,'' useful information gained, perceptions of agent personas, and the utility of different system features. We asked specific questions about which features participants found most helpful for critical thinking and exploring different perspectives and comparison between the two conditions. The exit interview script can be found in \cref{apdx:exit-interview-script}.

\subsection{Data Analysis}

To evaluate our research questions, we collected and analyzed a combination of qualitative and quantitative data, including system logs, think-aloud protocols, surveys, and exit interviews.

\subsubsection{Analysis of System Logs}
% \ta{I might cut this a lot and just say you logged all interactions with the system then say how you coded interactions}
We collected and analyzed all users' interactions with the system.
Specifically for users' textual interactions with agents (replies in \systemName\ and chat messages in group-chat condition), two researchers performed a qualitative analysis of users' inputs to the systems during these interactions. We analyzed users' input when replying to agents' posts during the forum condition, and the free-text input when interacting with agents in the baseline chat condition. We first coded these inputs to identify patterns of interaction, focusing on users' intents. Two researchers independently coded the data, and then met to discuss and resolve any discrepancies. the final code includes 13 categories, including \texttt{design}, \texttt{method}, \texttt{critique}, \texttt{expand}, \texttt{data-seek}, \texttt{reflect}, \texttt{risk}, \texttt{alternative}, \texttt{apply}, \texttt{compare}, \texttt{clarify}, \texttt{ethics/impact}, and \texttt{summarize}. Detailed descriptions of these categories are provided in Appendix~\ref{apdx:interaction-codebook}. 
Two researchers independently coded all user inputs using the codebook, achieving a Cohen's kappa of 0.88, which indicates strong inter-rater reliability. Discrepancies were later resolved through discussion with both researchers to reach consensus on the final coding results.
\label{sec:reply-coding}

\subsubsection{Analysis of Proposal Data}
\label{section:llm-judge-eval-proposal}
We also perform an LLM-as-a-judge evaluation of the quality of participants' research proposals. We use OpenAI's GPT-5 model to assess the proposal quality based on the same criteria as the self-assessment survey (i.e., coverage, significance, depth, feasibility, and clarity). We excluded relevance as a dimension since the relevance of the proposal and a user's actual intended research topic can only be self-assessed.
We prompt the model to rate each proposal on a 1-7 Likert scale across all six dimensions in a pairwise manner, by providing both the initial proposal and the revised final proposal as the context of the prompt. Then we calculate the difference in scores between the two proposals to measure the degree of improvement in proposal quality, where the delta for each metric \(m\) and pair \(i\) is \(\Delta_{i,m} = \mu_{i,m}^\text{final} - \mu_{i,m}^\text{init}\).
To ensure the consistency of the evaluation, we ran 10 independent judgments and averaged the numeric scores by each dimension (for each user and condition). 
To ensure the reliability of LLM-as-a-judge ratings, we conducted human validation by sampling 10 proposal pairs (5 with the highest LLM ratings and 5 with the lowest LLM ratings) and had two researchers rate them independently using the same rubric. 
We obtained a good inter-rater reliability across the five dimensions of an average Krippendorff’s alpha of 0.75, indicating substantial agreement.

\subsubsection{Analysis of Think-Aloud Data}
During the study sessions, we collected think-aloud data to gain insight into participants' cognitive processes. The analysis of this data focused on identifying moments of critical thinking, such as behaviors that align with Bloom's taxonomy, instances where participants identified contradictions or gaps in arguments, and ``aha'' moments indicating a change in perspective.

\subsubsection{Survey Data}
Post-session surveys were used to collect participants' subjective feedback on their experience. The surveys included established instruments such as the NASA-TLX for cognitive load and the System Usability Scale (SUS) for usability. We also included custom scales to measure perceived interdisciplinary clarity, adapted from prior work~\cite{stokols2015transdisciplinary, owan2024development}, which covered conceptual, methodological, role, and communication clarity. Participants also completed a self-assessment of their critical thinking and reflection, covering a range of cognitive activities from knowledge recall to self-regulation. Finally, a self-assessment of the quality of their research proposal was collected, evaluating aspects such as coverage, significance, relevance, depth, feasibility, and clarity.

\subsubsection{Exit Interview}
We conducted 15-minute semi-structured exit interviews to gather qualitative feedback on the user experience. Questions were designed to elicit reflections on moments that were particularly helpful or challenging for their critical thinking process and to identify which specific features supported the exploration of different perspectives. The exit interview script can be found in~\cref{apdx:exit-interview-script}.

\section{Findings}

To understand how \systemName\ supports users' interdisciplinary deliberation, we conducted a mixed-methods analysis of user interactions, survey responses, and think-aloud data. Our findings aim to address the following research questions:
\begin{itemize}
    \item \textbf{RQ1:} \textit{How does \systemName\ impact users' proposal revisions and quality?}
    \item \textbf{RQ2:} \textit{How do users leverage \systemName's unique wfeatures?}    
    \item \textbf{RQ3:} \textit{How does \systemName's interaction design influence users' critical thinking activities?}
\end{itemize}

% \yiren{Show cases/examples of agent and human deliberation; use screenshots for validity}

% \yun{5.1 Survey + think-aloud to explain their perceptions.}

% \yun{5.2 Significantly Improved Discussion Outcomes  e.g., proposal quality; Critical thinking activities more and differently: figure 8.} 
% \yiren{combine qualitative think-aloud with proposal edit analysis}

% \yun{5.3 case studies (compare multi-agent vs single-agent (remove threaded)}
% \yiren{Case studies of behavior funnel/sequences}

% \yun{5.1 different critical thinking involved, examples, 5.2 compare overall writing quality. 5.3 no different cognitive load etc.}

\subsection{ \systemName\ Enhances  Proposal Quality without Increasing Cognitive Load (RQ1)}
In this section, we first present improvement of users' proposal edits across each condition; we then present two case studies to demonstrate typical usage patterns of the two systems; finally, we present the user-perceived usability of \systemName\ and each feature.

\subsubsection{Improved Clarity and Feasibility of Written Proposals}
The \systemName\ condition led to better improvements in proposal quality compared to the baseline condition, as evaluated by both LLM-based assessment and user self-assessment.
% We also conducted an LLM-based evaluation of the proposal quality, using OpenAI o3 to assess the proposal quality based on the same criteria as the self-assessment survey. Detailed method of the LLM-as-a-judge evaluation can be found in \cref{section:llm-judge-eval-proposal}.
% We then compared the proposal quality scores yielded from the LLM-as-a-judge process between the forum-based and group-based conditions (averaged at the user level). \yiren{move all method to method section}
The forum condition showed larger mean gains than chat on the dimensions of \textit{Clarity} (M=0.87 vs. 0.39; $t=-2.15$, $p=.039^{*}$) and \textit{Feasibility} (M=0.56 vs. 0.23; $t=-2.37$, $p=.024^{*}$). 
The \textit{Overall} rating also favored forum condition but not significantly (M=0.90 vs. 0.66; $p=.235$). 
% \textit{Coverage}, \textit{Depth}, and \textit{Significance/Novelty} also shared this trend but differences were non-significant.
% \textbf{Comparison of proposal quality self-assessment Survey rating pre- and post-session.} 
Users' self-perceptions mirrored these results, where participants gave higher rating improvement (comparing pre- and post-session ratings) of proposal quality under \systemName\ compared to the group-chat condition, with the largest gains in Coverage (M=0.41 vs. M=0.19) and Significance (M=0.50 vs. M=0.25).

\subsubsection{Significantly More Revisions and Better Motivated Proposals} 
The use of \systemName\ also led to significantly more revisions overall (M=5.35 vs. M=2.19). A chi-squared test revealed a strong association between the condition and which fields of the notepad users edited ($\chi^2 = 180.33, p < .001^{***}$). Post-hoc tests~\footnote{Two-proportion z-tests with Benjamini-Hochberg correction.} showed that forum users edited the \textit{Motivation} section significantly more than chat users (33.1\% of revisions vs. 4.3\%; $z=-3.98$, $p<0.001^{***}$). While other per-field edit counts and magnitudes showed similar trends favoring the forum condition, they were not significant after correction. 
It is worth noting that we found no significant differences in the amount of revisions over the \textit{Notes}, where most of the additions were made using the quick note-taking feature (as described in \cref{sec:system-design}). This observation suggests that although users find a similar amount of relevant content in each condition, they are more likely to translate this content into structured proposal changes during the use of \systemName. One example of this was T18, whose proposal edits using \systemName\ showcased how they were able to utilize diverse agents' inputs to strengthen different aspects of their initial research idea, as shown in \cref{fig:t18-side-by-side}.

% It is worth noting that, T18's proposal edits during the forum condition involves stronger thinking and reasoning from themselves, whereas during the group chat the edits are mostly copy pasting~\cref{apdx:t18-proposal-edit-full}. 
% When participants were asked to reflect on their rationale of proposal edits, ...

%TC:ignore
\begin{figure*}[!ht]
\centering

\begin{tcbraster}[
  raster columns=2,
  raster equal height,
  % raster left skip=0pt,
  % raster right skip=0pt,
  % raster column skip=6mm
]

% ===== Left box =====
\begin{tcolorbox}[
  enhanced,
  % breakable,
  colback=white,
  colframe=black!40,
  title={\normalsize T18's proposal edits (\textsc{\systemName}):\\ \emph{Family Conversational Agents}},
  fonttitle=\bfseries,
  sharp corners
]
\footnotesize
% \difflegend
\medskip
\diffctx{Motivation:}
\diffnochg{I want to learn more about how people work with or receive service from a robot team instead of just one robot}\\
\diffadd{clarify aspects to explore - (human perception): efficiency, usability, and emotional comfort.}\\
\diffadd{clarify robot team attributes to explore - (robot team attributes): specialized and clearly differentiated roles versus homogeneous teams, and whether and how they show those attributes to the users (visibility of specialization, or social cues)}\\
\diffadd{Other factors that can affect the perception of the interaction: task complexity can vary the perception between the same user and the same robot team, also task nature}
\medskip
\diffctx{Related Work:}
\diffnochg{much prior exploration is about one human interacting with one robot, but not much about one human with multiple ones and how they feel about the team dynamics}\\
\diffadd{transparency related work is limited in multi-robot setting}\\
\diffadd{some current techniques can help increase transparency}\\
\diffadd{too much competency can lead to different perception}
\medskip
\diffctx{Methods:}
\diffnochg{literature review, select a domain}\\
\diffadd{potentially healthcare service}\\
\diffnochg{, select a scenario for detailed experiments, user study design}
\medskip
\diffctx{Potential Outcomes:}
\diffnochg{robot teammates' interaction quality will affect people's trust; in different cases, when robots make mistakes, people may have different tolerance to the mistakes}\\
\diffadd{now focus on add transparency during the process and control other variables like task complexity, but mention in discussion}
\end{tcolorbox}
% ===== Right box =====
\begin{tcolorbox}[
  enhanced,
  % breakable,
  colback=white,
  colframe=black!40,
  title={\normalsize T18's proposal edits (Group Chat):\\\emph{Human--Robot Interaction}},
  fonttitle=\bfseries,
  sharp corners
]
\footnotesize
% \difflegend
\medskip
\diffctx{Motivation:}
\diffnochg{how to improve transparency of human-robot-interaction specifically in a human - multiple robots settings}
\medskip
\diffctx{Related Work:}
\diffnochg{human robot one on one interaction but not much about robot team}\\
\diffagent{definition: transparency in robot-team collaboration is fundamentally ...}\\
\diffadd{social perspective related work: SAT}\\
\diffagent{effective explanation interfaces must support ``drill-down'' capability ...}\\
\diffagent{Another critical challenge as teams scale is maintaining mutual awareness ...}\\
\diffagent{adaptive transparency At a sociotechnical level ...}
\medskip
\diffctx{Methods:}
\diffnochg{literature rievew, tool design, propotype development, evaluation}\\
\diffagent{system design: expose their internal decision-making processes ...}\\
\diffadd{open question: strategies to balance detail vs overload?}\\
\diffagent{From the qualitative side, I'd suggest deploying adaptive transparency in a hospital logistics context ...}\\
\diffagent{That's a fantastic take—I'd echo the critical value of ``soft'' outcome measures ...}
\medskip
\diffctx{Potential Outcomes:}
\diffnochg{more efforts should be put on how to show the coordination between the robot teammates, and level of this transparency should be carefully considered}
\end{tcolorbox}
\end{tcbraster}
\caption{
Side-by-side comparison of T18's proposal edits across two conditions. Purple text with a leading [COPIED] refers to text chunk copied directly from agents' quotes and pasted in proposal (we omitted the full texts of these quotes except the leading sentences due to their often verbosity, full text can be found in \cref{apdx:t18-proposal-edit-full}). As shown in this figure, T18's edits using \systemName\ reflect more active thinking, whereas the edits using the group-chat interface are mostly copy-and-pastes.
}
\label{fig:t18-side-by-side}
\Description{Side-by-side tcolor boxes compare T18’s proposal edits made with the system (left, “Family Conversational Agents”) versus a group-chat interface (right, “Human–Robot Interaction”). Both columns are organized by sections—Motivation, Related Work, Methods, Potential Outcomes. Left (system): Edits synthesize and refine ideas. Motivation specifies dimensions to study in multi-robot teams (efficiency, usability, emotional comfort) and team attributes (specialized vs. homogeneous roles, visibility of specialization/social cues). Notes other factors (task complexity, task nature). Related Work highlights limited transparency research in multi-robot settings, mentions techniques to increase transparency, and cautions that high competency can shift perceptions. Methods: literature review, domain selection (potentially healthcare), scenario selection, user-study design. Outcomes: trust varies with interaction quality; tolerance of mistakes depends on context; plan to add transparency while controlling variables like task complexity. Right (group-chat): Edits are broader and include many pasted agent quotes marked [COPIED] (purple). Motivation: improve transparency in human–multi-robot interaction. Related Work: mostly one-to-one HRI; copied notes define transparency, stress drill-down explanations, mutual awareness, and “adaptive transparency.” Methods: literature review, tool/prototype and evaluation; copied suggestions to expose decision processes, balance detail vs. overload, and try a hospital-logistics deployment; emphasize soft outcome measures. Outcomes: show teammate coordination; tune transparency level.}
\end{figure*}
%TC:endignore

% Participants rated their user experience, cognitive load, and the helpfulness of system features in a post-session survey. 
\subsubsection{No Major Differences of Cognitive Load between \systemName\ and Control} 
\label{subsec:T17Rq1}
While overall UX and usability ratings (7-point Likert scale) were high and did not differ significantly between the \systemName\ and group-based conditions, the forum interface was perceived as slightly less demanding. Regarding cognitive load, \systemName\ was rated as requiring slightly less ``Mental Demand'' (M=3.89 vs. 3.94), ``Effort'' (M=3.72 vs. 4.13), and inducing less ``Stress'' (M=2.94 vs. 3.50) compared to the group-chat interface, though these differences were not statistically significant. More detailed results can be found in \cref{apdx:post-survey-ratings}.
Participants found the persona-based agents more helpful in the forum condition (M=5.67, SD=1.08) than in the group-chat condition (M=5.00, SD=1.37). For the forum interface, the most valued features were the \textit{Forum-based Reply Interaction} (M=5.76, SD=1.48), \textit{Expert Persona Customization} (M=5.67, SD=1.08), and the \textit{Forum-based Layout} (M=5.39, SD=1.29).
Across participants' think-aloud data, we found that participants valued reading multiple expert viewpoints in dialogue, rather than a single stream of answers from a single persona (N=9). 
As noted by T10, \userquote{... this interface makes more sense than down here (the group-chat interface), just because ... I think it's just like easier to understand what everyone is adding to the conversation.}
T17 mentioned preference for the broader coverage from different perspectives, \userquote{I like the first one, because there is discussion between different roles}, and mentioned being \userquote{... inspired by some perspectives from several agents ...}. 
Participants also noted that agents \userquote{respond[ing] to each other} made threads clearer (T17) and, when combined with the mind-map, helped them \userquote{mentally group what was going on} across questions (T6).

\subsection{ \systemName\ Achieves Design Goals and Catalyzes New Affordances 
(RQ2)}
In this section, we analyzed users' interaction system logs and their think-aloud data to uncover common affordances, how they interpreted the system and agents, and new observations about user behaviors and feature use case scenarios emerged beyond the initial design goals.

%\subsubsection{Supporting Targeted Deliberation and Verification via @-mention and reply}

%\subsubsection{Effective use of @-mention and reply: single agent for requesting specialized knowledge and multi-agent for gathering critical feedback and evaluation for specific aspect of research (e.g., method) (DG1)}

\subsubsection{Designed Affordances of @-mention and reply: Panels, Threads, and Expert Targeting (DG1)}
In general, we found that users actively sought to synthesize information from multiple disciplines, often using @-mentions to bring different expert agents into a single conversation thread. Users were able to utilize the @-mentions effectively (4.67 times per user on average). We also found that forum replies revealed that 45.1\% (23 out of 51) of user replies were interdisciplinary. The proportion of interdisciplinary replies increased when users explicitly used @-mentions, where 58.3\% of such replies were cross-disciplinary, compared to 41.0\% for replies without mentions. %, where the user's query engaged a different domain from the parent post's agent

\begin{figure}[ht!]
    \centering
    \includegraphics[width=1\linewidth]{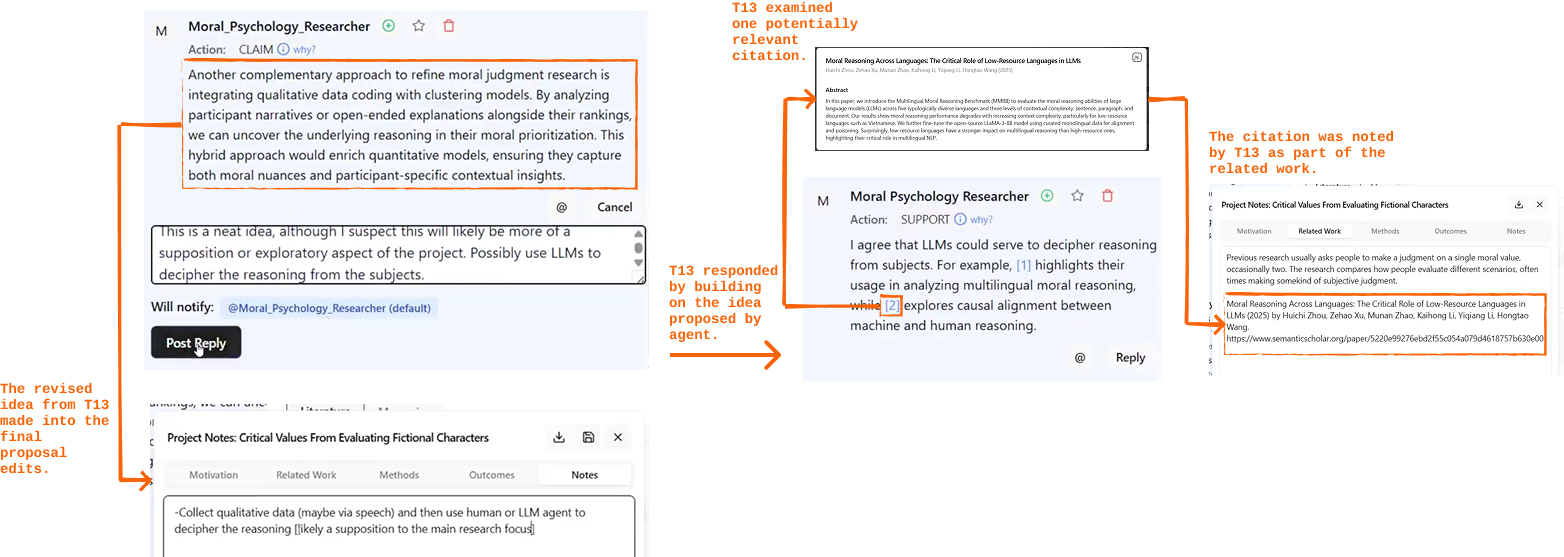}
    \caption{An illustration of how T13 interacted with ``Moral Psychology Researcher'' agent and collaboratively developed an idea through user reply; the screenshot of the notepad shows the immediate edit made by the user after reflecting on the agent's response.}
    \label{fig:case-study-T13-single-agent}
    \Description{
    Figure showing an annotated workflow across screenshots. Left panel: A comment from ‘Moral_Psychology_Researcher’ proposes integrating qualitative data coding with clustering to analyze participants’ narratives and rankings in moral‐judgment studies; the reply area is visible. A margin note reads ‘The revised idea from T13 made into the final proposal edits.’ Right, top: T13 opens a paper titled ‘Moral Reasoning Across Languages: The Critical Role of Low-Resource Languages in LLMs.’ Right, middle: A reply from ‘Moral Psychology Researcher’ supports the agent’s idea and cites items [1] and [2]. A label says ‘T13 examined one potentially relevant citation.’ Right, bottom: The citation is added under a ‘Related Work’ tab in project notes with the paper’s title and link. A label notes ‘The citation was noted by T13 as part of the related work.’ The annotations emphasize how an agent’s suggestion was validated with literature and incorporated into the final proposal.
    }
\end{figure}

For example, a case study of T13 is shown in \cref{fig:case-study-T13-single-agent}, a behavioral economy researcher on discussing the impact of AI on human decision-making. During exploration, T13 came across a ``Moral Psychology Researcher'' agent who proposed the idea of combining qualitative data coding with clustering models. As the agent explored the methodological extension, T13 explicitly endorsed the idea through reply and reflection, and then further tightened the scope to include deciphering of subject reasoning using LLM. The idea and its retrieved citation were later recorded in the proposal.

Other participants also demonstrated similar use of the reply feature to interact with single agents. More specifically, users would employ single @-mentions to request deep, specialized knowledge from a specific expert. 
One common use by participants was to seek concrete answers to questions that required domain-specific knowledge, such as asking a ``@HCI Researcher'' about user evaluation study design (T1) or an ``@Healthcare Policy Analyst'' about examples of infrastructure gaps and cultural barriers in engaging caregivers in rural areas (T11).

\subsubsection{Eliciting Feedback from Multiple Agents via Sensemaking Structured Dialogue Map (DG2)}
% disagree -- debate 5.4.1 
In general, participants appreciated that agents under the forum condition provided more critiques and disagreement compared to the group-chat condition and common off-the-shelf chatting applications and perceived them as beneficial to their ideation process (N=6). This is largely due to \systemName's feature that enables visualization of agents' deliberation action and rationales.

\begin{figure}
    \centering
    \includegraphics[width=1\linewidth]{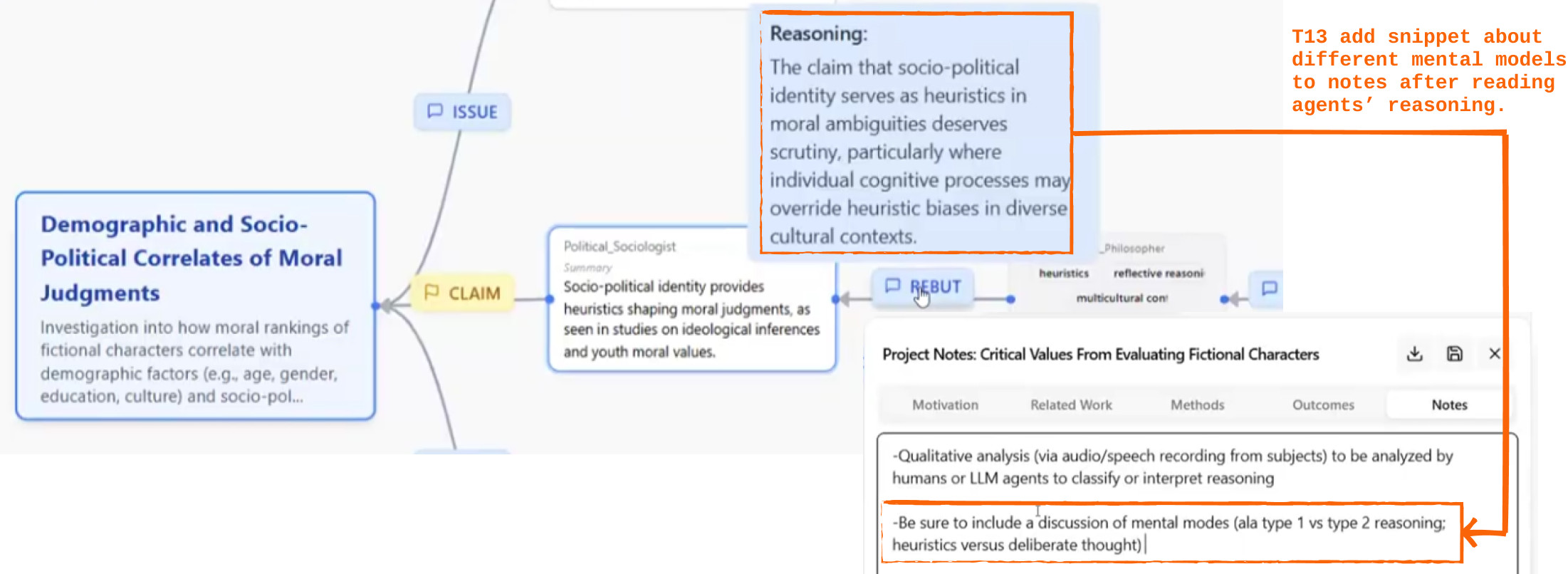}
    \caption{An illustration of how T13 used rationale between agent action for sensemaking of agent responses; the reasoning of the ``Rebut'' action by the ``Moral Philosopher'' inspired users' consideration about different cognitive mental models.}
    \label{fig:T13-mindmap-rationale}
    \Description{
    Figure showing an annotated workflow. Mind-map panel with nodes for ISSUE and CLAIM linked to ‘Demographic and Socio-Political Correlates of Moral Judgments.’ A REBUT button is visible. A ‘Reasoning’ pop-over argues that treating socio-political identity as a heuristic in moral ambiguity needs scrutiny, noting individual cognition may override biases. Project Notes panel (Notes tab) where T13 adds a bullet reminding to discuss different mental models—Type 1 vs Type 2 reasoning, heuristics versus deliberate thought.
    }
\end{figure}

In the case of T13, they switched to explore a different thread by using the mind map feature. They skimmed the sentence-level summaries of agents' responses, and decided to dig deeper into a branch related to socio-political identity shaping moral judgments. The participant utilized the rationales of agent actions (displayed when the cursor hovers over an action label) to help them understand the thinking process behind agents' responses. As shown in \cref{fig:T13-mindmap-rationale}, T13 took note of one of the agents' rationale of a ``REBUT'' action, by combining it with their own ideas.
% \yiren{Find more examples linking actions between agent @s}
Another example is T5, who was prompted to ask follow-up questions after seeing a reply with action type ``QUESTION'' from ``Cybersecurity Specialist.'' This reply was asked in context to another response from ``HCI Research'' giving an opinion on designing for privacy preservation during real-time feedback loops. ``Cybersecurity Specialist'' asked a two-part question to this response: 1) how to ensure user consent, and 2) what are practical techniques that can be applied for this use case. T5 then commented on his rationale for involving multiple agents, noting that since it was both a \userquote{privacy question... [and] an engineering question}, they needed to tag both \userquote{the policy ones... and the engineers.} They then included five agents (``@AI Researcher,'' ``@Data Science Ethicist,'' ``@Privacy Advocate,'' ``@Machine Learning Engineer,'' and ``@Ethics and Policy Researcher'') to form a discussion over the topic of blockchain-based encrypted consent workflow for an ML pipeline.

T13 also adopted agents' suggestions within a self-chosen scope, with the goal of protecting the core contribution being original and validated.
This discussion was later incorporated in their final proposal: 
\begin{displayquote}
\textit{
Collect qualitative data (maybe via speech) and then use human or LLM agent to decipher the reasoning \textcolor{darkgreen}{+ likely a supposition to the main research focus}.
\textcolor{darkgreen}{+ Qualitative analysis (via audio/speech recording from subjects) to be analyzed by humans or LLM agents to classify or interpret reasoning}.
\textcolor{darkgreen}{+ Key Terms to Check — Chain of Thought (CoT)}
\textcolor{darkgreen}{+ Be sure to include a discussion of mental modes (ala type 1 vs type 2 reasoning; heuristics versus deliberate thought).}
} --- addition over the initial method section of the proposal by T13
\end{displayquote}

This preference for critical discussions is also reflected upon by many other participants, as mentioned by T9 \userquote{if you're just going to all keep agreeing... I didn't really learn anything new}, noting that LLMs should \userquote{give us different view[s]... another way of seeing it} beyond what papers already provide (T3). 
The visible critique across personas normalized constructive disagreement, as mentioned by T17 \userquote{I like the debating a lot ... criticism make projects better}, and created opportunities for new perspectives, similar to lab meetings where colleagues articulate why they hold a view and propose ways forward. 
%% active thinking and intepretation of agents' thinking process
T7 noted that \systemName\ drove them to engage in more active thinking, \userquote{I need to read them and ... and use my own analysis to keep the discussion going ... ,} while in the group-chat condition, the participant expressed a lack of trust as most content is solely generated by agents without much human input.
Several participants also sought to actively steer divergence, \userquote{choose [an agent] to take an agree path versus a disagree path versus bringing up a new question} (T7), and valued getting different opinions beyond own research domains, as mentioned in \userquote{... [as biology researchers] we never think about ethics, laws, or anything like economics... } (T8). This preference reflects the need to \userquote{hear what other people are thinking} (T6) and bring a \userquote{fresh set of eyes} to their work (T13). T9 also noted the convenience of being able to \userquote{talk to anyone} when needed.
As T6 noted, \userquote{this simulates like how I interact with my lab mates ... we're not just pulling up papers ... we're just thinking about the problem itself, like, how would we go about tackling it. What methods would we use?} and \userquote{I just want to hear what other people are thinking.}

\begin{figure}[ht!]
    \centering
    \includegraphics[width=1\linewidth]{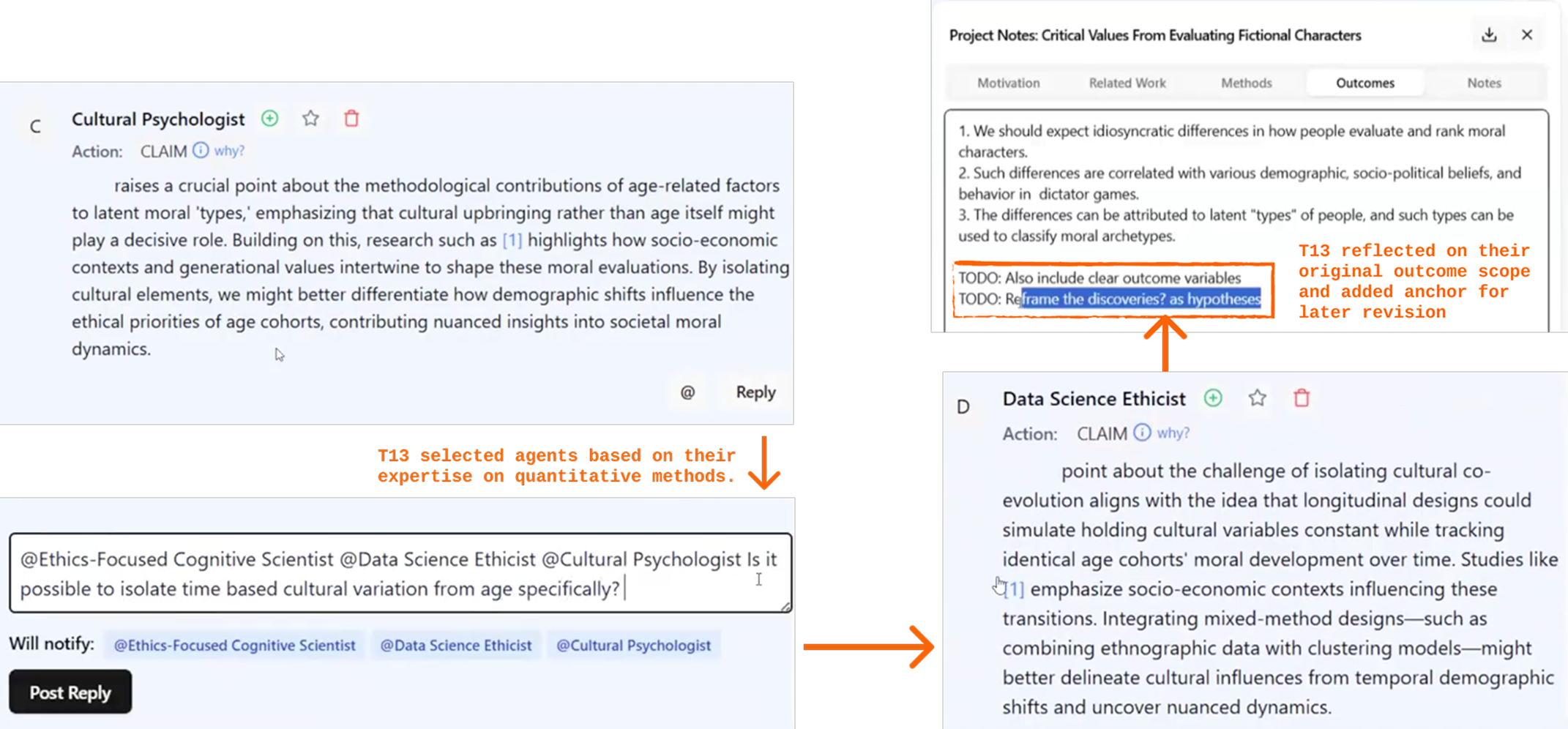}
    \caption{An illustration of how T13 involves multiple agents using the @-mention feature to gather input from different perspectives about a methodological question; the participant rationalized the selection of agents based on their expertise in quantitative methods.
    }
    \label{fig:T13-multi-agent-remix}
    \Description{
    Figure depicting T13’s reasoning loop across multiple UI panels. Upper left: Agent ‘Cultural Psychologist’ issues a CLAIM: cultural upbringing and socio-economic context may better explain latent moral ‘types’ than age alone; isolating cultural elements could clarify cohort-level ethical priorities. Lower left: T13 composes a reply tagging @Ethics-Focused Cognitive Scientist, @Data Science Ethicist, and @Cultural Psychologist, asking whether time-based cultural variation can be isolated from age effects. A note reads, ‘T13 selected agents based on their quantitative methods.’ Upper right: Project Notes → Outcomes lists expectations; T13 inserts TODOs: ‘Also include clear outcome variables’ and ‘Reframe the discoveries as hypotheses,’ noted as anchors for later revision. Lower right: Agent ‘Data Science Ethicist’ posts a CLAIM highlighting the difficulty of isolating culture, proposing longitudinal designs and mixed methods (e.g., ethnography with clustering models) to hold cultural variables constant while tracking age-cohort changes and uncover nuanced dynamics.
    }
\end{figure}

\subsubsection{Emergent Affordances: Proactive Engagement and  Verification with  User-generated TODOs} 

While T13 actively involved multiple agents using the @-mention feature to gather input from different perspectives about a methodological question, as shown in \cref{fig:T13-multi-agent-remix}, we observed an unexpected workflow. 
Specifically, in contrast with interacting with one agent to gather details about related work (example of T13 in \cref{fig:case-study-T13-single-agent}), the user planned TODO lists using a notepad, and called in multiple users to tackle one TODO at a time, gathering critical feedback about method design.
T13 decided to engage with multiple additional agents with different backgrounds, as they found the initial response generated by ``Cultural Psychologist'' lacked the desired details about the proposed method.
As T13 noted\userquote{Well, how would you (differentiate the demographics based on cultural elements)? ... let's ask one of the quantitative people ... ,} they selected ``Ethics-Focused Cognitive Scientist,'' ``Data Science Ethist'' and ``Cultural Psychologist'' agents to respond to the question, which later converted into anchors in the outcome section for later follow-up.
% \userquote{Is it possible to isolate time based cultural variation from age specifically? In principle, we want to study effectively the same sample of people at age 20 versus age 50, while holding cultural aspects constant. This seems impossible given that culture is always co-evolving.} --- T13
This case also surfaced an emergent affordance: T13 left ``TODO'' anchors as self-assigned checkpoints for later verification. These anchors flagged agent-suggested claims for subsequent verification (e.g., double-checking on citations). This reflects the user's proactive engagement in thinking and an intention to keep the core contribution original.

When agents conversed with one another, participants found it helpful to observe the exchange of agents acknowledging concerns and offering potential solutions (\userquote{acknowledging the fact that that is a concern, and then also providing solutions to mitigate or overcome that}) helped them evaluate trade-offs and refine questions. 
As T6 noted, one agent may surface an inferred stance, \userquote{users might also expect an assistant ... to infer ... the system ... would ... convey some sort of interpretation or stance}, and that \userquote{the next agent is then acknowledging the fact that that is a concern, and then also providing ... solutions}, which enabled them to \userquote{pick up their brains and come up with my own reasoning ... think about it in this new light.}
These interpretations suggest the system's feature that allows users to participate while the agents build on each other's thoughts facilitates users' active thinking and reflection, and helps them build upon agents' reasoning and thinking process, not through mere recall of information but critical reflections.

\subsection{\systemName\ Facilitates Higher-Order Critical-Thinking Activities (RQ3)}

\subsubsection{Commonly Applied Workflows: Charting and Synthesizing Problem Spaces via Branching Multi-Agent Deliberation}

We observed that participants often utilized the persona-based agents not only to contextualize their research ideas in unfamiliar domains but also to understand the thinking processes behind different expert roles (N=7). More specifically, participants found the agents helpful for exploring areas beyond their own expertise.

An example of this is T17, who developed a strategy of decomposing complex questions into specific queries targeting chosen experts. In T17's words \userquote{For the 1st tool I ... mentioned the specific agent to answer my question ... break the big questions into small questions.} This strategy was supported by the \systemName\ when compared to the baseline condition, as noted by T17: \userquote{But for the second tool (the group-based chat interface) I cannot do that. I can just talking with several agents at the same time ... I cannot mention, specify this agent I want to talk to, and there are too much information.}

\begin{figure}
    \centering
    \includegraphics[width=1\linewidth]{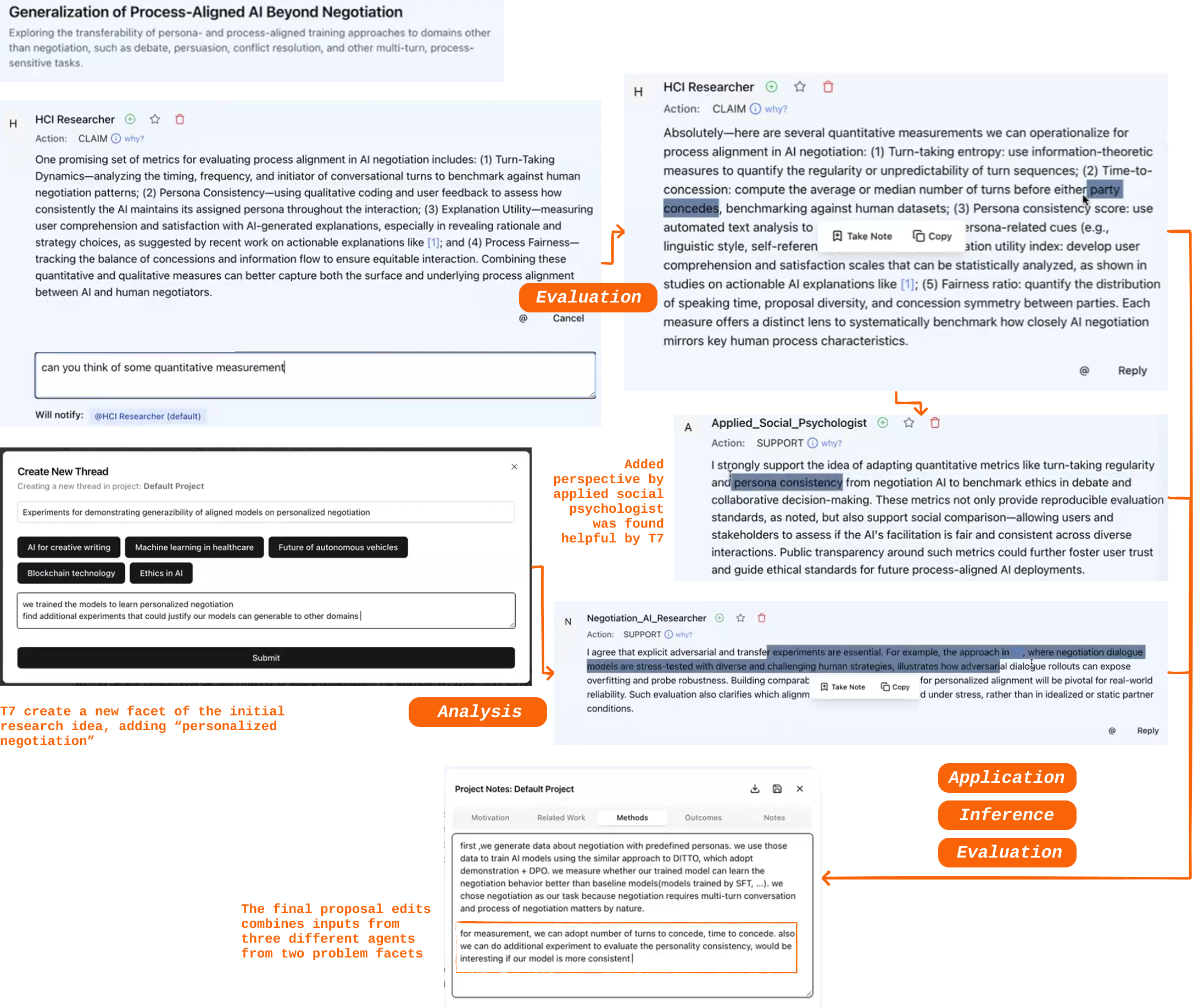}
    \caption{An illustration of T17's workflow, showcasing how the participant decomposes the initial research idea into two different facets during exploration: 1) process-aligned evaluation and 2) personalized negotiation. The user utilized the ``create new thread'' feature to initiate a new discussion about the second facet; the final proposal edit combines inputs from three different agents throughout the two facets. 
    }
    \label{fig:T17-decompose}
    \Description{
    Figure summarizing a collaborative design loop for evaluating process-aligned AI beyond negotiation. A header shows the project title. Top-left panel: an HCI Researcher posts a CLAIM proposing four metric areas—turn-taking dynamics, persona consistency, explanation utility, and fairness-related scores. A prompt asks for ‘quantitative measurements.’ Top-right panel: another HCI Researcher enumerates measurable metrics: turn-taking entropy; time-to-concession (average/median turns before first concession, benchmarked on human data); persona-consistency scores with automated text analysis; user-study scales for explanation/comprehension; and fairness ratios of speaking time, proposal diversity, and concession symmetry. Right column: an Applied Social Psychologist strongly supports adapting these metrics for benchmarking and transparency. A Negotiation-AI Researcher supports adversarial/stress-test experiments and real-world personalization. Center-left panel: a ‘Create New Thread’ dialog introduces a new facet, ‘personalized negotiation.’ Bottom-left panel: Project Notes (Methods) describe collecting negotiation data with/without personalized personas and measuring persona consistency and turns-to-concede, noting added experiments; a caption states that final edits combine inputs from three agents across two problem facets. Orange callouts label stages: Evaluation, Analysis, Application, Inference, Evaluation.
    }
\end{figure}

The participant then interacted with agents to scaffold understanding. T17 reflected: \userquote{In real world research discussion, like, different people have different backgrounds, knowledge. And we can gain a lot of like visions knowledge from other people's. And in this tool I kind of feel similar because I feel like I... was inspired by some... perspectives from several agents.} The reply-to threading did reinforce the participant's perceived clarity of the collaborative discourse: \userquote{Agents respond to each other makes the ... thread very clear ... it is very clear to review those comments.} 

\systemName's agent deliberation then helped shape T17's proposal from a broad idea into a more feasible study. After agents mentioned the need for ``more metrics'' and T17 observed that ``quantitative measurement would be better,'' they added concrete measures like ``number of turns to concede'' and ``time to concede,'' reflecting \textit{Evaluation} and \textit{Inference}. 
Discussion about personas' ``different strategies of negotiation'' helped T17 highlight negotiation as a multi-turn process where the process itself matters. Mentions of ``consistency'' further motivated an added experiment on personality consistency, reflecting \textit{Application} and \textit{Inference}.
% \textcolor{grey}{\sout{analyses}}
As shown in T17's note edits from \cref{fig:T17-decompose}, it is evident how agents' suggestions made into the final edited proposal. 

Other participants also found agents useful for contextualizing their ideas in unfamiliar domains.
One noted the system helped them \userquote{think about applications beyond the usual [research topic] scenarios} by providing concrete examples (T18). For discussions involving multiple agents, another participant found it helpful to \userquote{very quickly see how different disciplines would go about solving the problem} (T16). Agents also provided context-specific reasoning, for instance, by explaining why a solution applicable in one region might fail in the US due to \userquote{infrastructure gaps} (T11), thus helping users ground their ideas in realistic settings.
Some participants went deeper and reflected on the reasoning process of the personas. T16, for example, noted that through the dialogue, they could see how an engineer persona thinks about success in terms of \userquote{what kinds of metrics you can use to measure your success... how do you concretely turn that into measurable indicators}. T16 found this insightful, stating, \userquote{I could understand more about... as that kind of persona, how their way of thinking changes... Through multi-round interactions between different agents, I can know how they approach thinking about a problem.} T9 wanted to take this further, as noted in \userquote{create my own pipeline... force each one of the agent to ask questions... And once they have the questions and answer, then I force them to rebute...}, suggesting the need to directly control agents' thinking process beyond mere interpretation.

\subsubsection{Varied Critical Thinking Activities with Multi-Agents}

\begin{figure}[htbp]
    \centering
    \begin{subfigure}[t]{1.0\linewidth}
        \includegraphics[width=\linewidth]{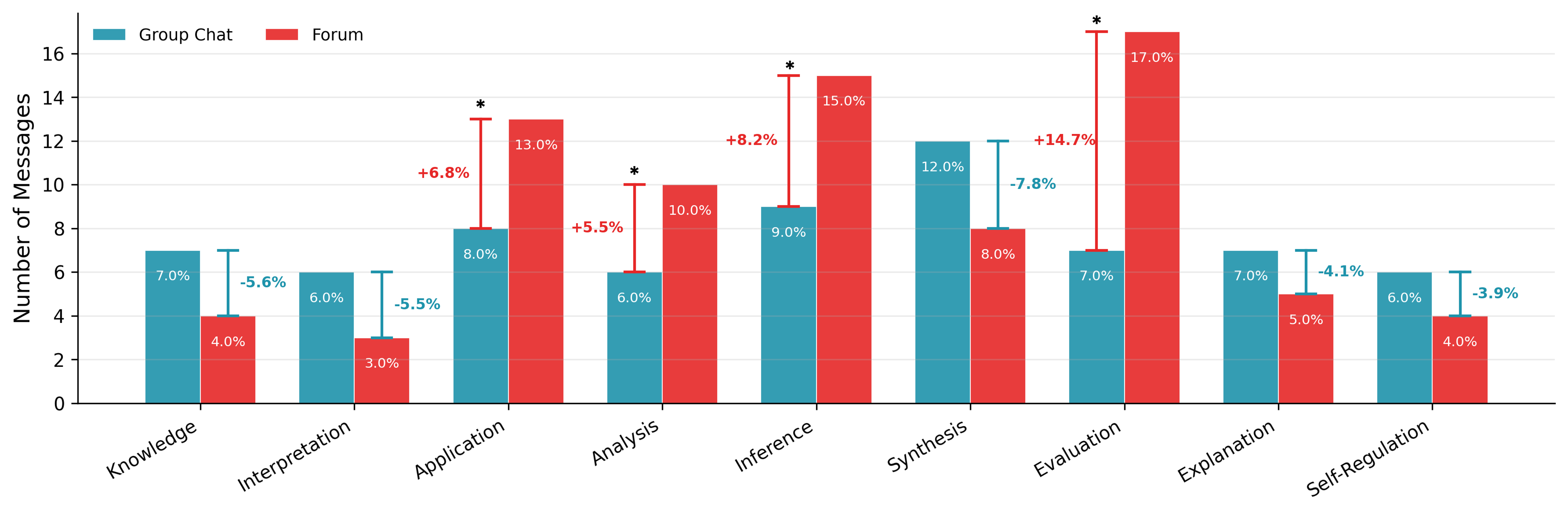}
        % \caption{Facione Skills Differences: Forum vs Chat (Positive = Forum Higher, Negative = Chat Higher). 
        % \yun{users' replies in two modes:  replies with different critical thinking activities. add total counts (significantly more amount) and \% (CHI-square to show distribution different). findings: give qualitative examples!! in the discussion, group certain behaviors into less or more critical thinking categories by citing references.}
        % }
        \label{fig:facione-difference-plot}
    \end{subfigure}
    % \hfill
    % \begin{subfigure}[t]{0.49\linewidth}
    %     \includegraphics[width=\linewidth]{imgs/facione_4_top_skills.png}
    %     \caption{Facione Skills Differences (Message Counts)}
    %     \label{fig:facione-difference-plot-co}
    % \end{subfigure}
    % \caption{Comparison of user-initiated messages coded with Facione's critical thinking skills across conditions.}
    \caption{Comparison of user-initiated messages coded with Facione's critical thinking skills across conditions (Positive = \systemName Higher, Negative = Group Chat Higher).
    % \yun{plase mark which ones are significant by adding a star on the top...}\yiren{added}
    }
    \label{fig:facione-plots-combined}
    \Description{
    Figure: Group Chat (teal) vs Forum (red) message distribution by critical-thinking facet. Y-axis = Number of Messages. For each facet, percentages within bars indicate share of all messages in that condition. Forum shows higher shares and taller bars for Application (13\% vs 8\%), Analysis (10\% vs 6\%), Inference (15\% vs 9\%), and Evaluation (17\% vs 7\%); these pairs are marked with asterisks to denote statistically significant differences. Group Chat leads on Knowledge (7\% vs 4\%), Interpretation (6\% vs 3\%), Synthesis (12\% vs 8\%), Explanation (7\% vs 5\%), and Self-Regulation (6\% vs 4\%). Error bars depict variability.
    }
\end{figure}

In order to obtain a more systematic understanding of how participants engage in critical thinking activities during system use,
We analyzed users' reply messages in \systemName\ and compared them with their chat messages in the group-based condition, by annotating the messages with the critical thinking codebook described in~\cref{sec:reply-coding}. 
Results show that \systemName\ elicited significantly more \textit{Method} (15.6\% vs. 4.8\%; $t=2.04$, $p=.044^{*}$) and \textit{Alternative} exploration (12.5\% vs. 1.6\%; $t=2.45$, $p=.017^{*}$) activities. Other categories (e.g., \textit{Risk} assessment: 10.9\% vs. 3.2\%, $p=.089$; \textit{Critique}: 10.9\% vs. 9.5\%, n.s.) trended in the same direction but did not reach statistical significance given the smaller message sample. As shown in Figure \ref{fig:facione-plots-combined}, the results reveal a significant difference in the distribution of critical thinking skill codes between the two conditions ($\chi^2=5.68, p < .05^{**}$), with users using \systemName\ demonstrated more \textit{Inference} (+8.2\%), \textit{Analysis} (+5.5\%), \textit{Application} (+6.8\%), and \textit{Evaluation} (+14.7\%) activities during interactions with agents. 

\subsubsection{\textit{Inference}, \textit{Application},  and \textit{Evaluation} Fostered by \systemName} 
While the relative frequency of \textit{Inference}, \textit{Application}, and \textit{Evaluation} codes was lower, skills related to communication and reflection were more prominent under the group-chat condition. 
Specifically, chat interactions contained more instances of \textit{Knowledge}, \textit{Interpretation}, \textit{Self-Regulation}, and \textit{Synthesis}. Users appeared to use the group-chat interaction for seeking clarification and expanding on ideas.

During the use of \systemName, as an example of \textit{Evaluation}, T8 asked, \userquote{but CRISPR based imaging has not been validated with high significance thus, would lead to nonsensical analysis} in \systemName, actively questioning the validity of the information provided by the agents, while most of T8's messages in the group-based condition are information-seeking questions. 
T16 also demonstrated \textit{Application} by reflecting on how to implement an idea in practice through requesting demonstration given a user-defined example: \userquote{In addition to interactive repair, are there any other design patterns that can support parent-child joint engagement? For example, how can the CAs initiate a conversation or an activity that involves both parties? @HCI Research do you know any implementations?} This question was directed towards ``Developmental Psychologist,'' who previously raised the idea of ``interactive co-repair'' in the discussion. The user grew intrigued by this point and proceed to inquire about other design patterns that could be applied in this context, also by tagging another agent to increase coverage.
\textit{Inference} behaviors are associated with envisioning potential risks and ethical concerns relevant to the proposed ideas. For instance, T5 reviewed the idea of integrating circuit theory into transcriptomics using graph-based methods, which proposed by ``Computational Biologist.'' T5 commented that although the idea was interesting, there might be potential issues based on their existing knowledge of this field, thus asking the question \userquote{Since the full gene regulatory network has not been explored, what can be the problems?} 

Participants' think-aloud data also revealed that they were more likely to engage in activities such as questioning validity, examining assumptions, and refining interpretations when using \systemName. More specifically, participants were more likely to ask questions about the validity of the information provided by the agents (e.g., \userquote{It should be geochemist, not environmental chemistry ... that would be more specific.} --- T10) and to reflect on their own understanding of the topic (e.g., \userquote{... this one (Cognitive Science Researcher) is very heavy on like the cognitive stuff. So I'm thinking that they maybe have some sort of background in Psychology or something like that ... } --- T6). 

\subsubsection{Complementary Engagement in the Group-Chat Condition for More Specialized Inquiries} 
When using the group-chat condition, the most common observations are information-seeking queries like \userquote{What coding assistants exist that I could test with using user studies?} (T7) and \userquote{Give me a literature review on algorithmic accountability in the past 5 years} (T9), which resemble typical off-the-shelf search engine and chatbot queries. 
For queries that fall under the \textit{Synthesis} category, participants tend to send out agent-led synthesis queries, which offload most of the reasoning to agents, such as \userquote{How could we frame a research proposal looking to understand the mechanisms by which we can increase research transparency between participants and researchers?} (T11) or even \userquote{Elaborate on the research gap and draft a research proposal?} (T9). This is coherent with our observations that they focus on specialized knowledge and gather more information about a given topic.

Users also commented on the complementary roles of \systemName\ and the group-chat design in supporting different aspects of critical thinking.
\systemName's panel-like interaction design to support critical thinking and reflections in a controllable manner that allows users to inject more of their own thinking and mitigate anchoring bias. During the exit interview, T9 explicitly compared \systemName\ with the STORM system~\cite{jiang2024into} with respect to \systemName's capability that allows users to perform more fine-grained control over discussion: \userquote{... it's very important to know where things go wrong so we can do counterfactual explanations and follow-up, and I think a pre-generated wall of text just doesn't do that...}, which helps the user to \userquote{ ... not be as biased to a preconceived notion ...} 
On the other hand, participants commented on the scenarios the group-chat is preferred for quick direct Q\&A about simpler information-seeking questions due to lower overhead, as noted by T6 \userquote{... (when using \systemName) before you get to that point there's a lot of information ... to take in and understand.}

\section{Discussion}
%%% evidence of user behavior in interacting with different agents:
% T5: Expanding to other similar generalization of solutions or problem settings 
% T9: contrasts it with a group chat setting where it feels easier to "talk to anyone" because "everyone is there."

%%% key takeaways from discussion:
% Visualize Deliberation: Users find value in agents that adopt critical, even conflicting, stances. This counters the typical agreeableness of LLMs and suggests that designing for ``adversarial discourse'' is a powerful way to stimulate more robust and creative ideation. Making agents’ deliberation visible (not just hidden CoT) elicited more critical thinking and proposal revision—and participants wanted stronger pushback—highlighting the value of designed adversarial discourse over sycophantic chat.
% Embrace ``Good Friction'': Intentionally designing for user control (e.g., requiring users to select agents) introduces beneficial ``friction.'' This friction slows users down, encouraging deeper engagement and more critical thinking compared to passively consuming a stream of AI-generated content.
% hybrid autonomy modes of interaction in-one-place: pair structured threads + a navigable mind-map with adjustable autonomy controls (re-engage agents, set stances, branch/summarize in place) to keep exploration and synthesis in one space and balance diversity with depth.
% \yiren{Yun: Organize the discussion into the following: 1) Interaction; 2) Outcomes; 3) Experience.}
In this section, we discuss major insights and takeaways from our findings by situating them in and extending discussions from prior research. 

\subsection{Beyond Dialogue: Enabling User-Steered Multi-Agent Deliberation}
% \yun{this title looks like the breath and depth in the old paper. How about: Beyond Dialogue: Advancing Critical Reasoning with \systemName} 
%% Structure (threading, visualization) are designed to support, not restrict, user steering.
The design of \systemName\ managed to scaffold discourse in ways a linear group-chat interface did not support. This led to improvement in discussion diversity without extra cognitive load, as observed in how participants used \systemName\ to support their own cognitive process: including decomposing complex problems into sub-questions (T17's case from \cref{subsec:T17Rq1}), ad-hoc ``panel-ization'' (RQ2), and contextualizing ideas across unfamiliar application settings (RQ3).
\systemName's design contributes to the growing body of research that explores design alternatives to traditional chat interfaces for LLM-based ideation systems that aim to balance between diversity and depth of exploration~\cite{liu2025personaflow,radensky2024scideator,jiang2024into,kim2025scaffolding}. Our findings provide further insights into multi-agent ideation support systems, in which case diversity vs. depth largely depends on the balance between user control and agent-driven self-orchestration~\cite{zheng2025deepresearcher,fourney2024magentic}. 
In \systemName's case, we used features such as @-mention and reply to provide users with more control over the exploration process. While it introduced friction for users wanting to quickly gather information and reach conclusions, it also encouraged slower, more deliberate thinking.

Past research has indicated the use of personas not only as a medium of information but also as a mechanism of interaction through persona customization~\cite{liu2025personaflow,sun2025persona} and selection of personas prior to debate~\cite{shi2024argumentative}. Our results suggest an alternative interaction design that allows users to ``panelize'' expert personas on the fly and branch into different threads. Participants demonstrated two valuable use case scenarios: 1) selecting and adding agents to better steer the exploration and consolidation of parallel threads (e.g., exploring decomposed sub-topics, and forming an ad-hoc panel of diverse background agents); 2) observing and interpreting the reasoning of agents for unfamiliar topics. 
These results suggest new approaches to using personas as interactions to advance multi-agent deliberation platforms. 
% \yun{I felt this section doesn't belong to adversarial multi-gagent deliberation. I'd suggest move it to design implications.}

%% When less-controlled agents could be complementary.. (consult with the group chat results).

\subsection{Better Sensemaking of Multi-Agent Deliberation through Visualization of Argumentation Acts}
\systemName's forum and map design allowing users to build a structure for deliberation among multiple agents. The proposed structure is built upon argumentation actions, including ISSUES, CLAIM, SUPPORT, REBUT, and QUESTION. Users select, combine, and re-engage specific agents to actively synthesize multiple perspectives rather than passively consuming information. Since active synthesis enhances users' intrinsic motivation and engagement~\cite{deci2012self}, this may explain why it leads to better ideation outcomes (RQ1) and drives active synthesis of knowledge and critical thinking activities (RQ3). 
One example of this is how T13 became intrigued with the rationale behind a ``REBUT'' action of an agent, and translated the rationale into a part of their final proposal. 
Theories in argumentation research~\cite{walton2007dialog} already posit that confronting opposing viewpoints strengthens reasoning, encourages sensemaking, and improves judgment quality. When people are challenged, they are more likely to scrutinize their claims, consider alternative perspectives, and refine their arguments, leading to deeper understanding and cognitive growth.
% While traditional adversarial discussion encourages active engagement and critical thinking, \systemName's design introduces explicit personas, each representing a distinct stakeholder or viewpoint. 
% \ta{not really clear how these points connect, suggest to remove 'While...performance' }
% \systemName's visualization of agents' reasoning processes provides insights and stimulate critical thinking beyond prior systems that merely use reasoning traces or chain-of-thought prompting to improve LLM performance~\cite{wei2022chain}.%, by demonstrating how visualizing the reasoning process of agents for human interpretation can further benefit users' critical thinking.
% \pao{Not sure how visualization is user-directed. If we are going to include this, the header is too narrow}\yiren{revised again to talk about adversarial deliberation} 
Furthermore, recent research~\cite{mei2025survey} strives to systematically formalize the methods to manage context (information payloads) in LLM-based systems as the rising area of ``Context Engineering.'' In this work, we contribute to the understanding of MAS context engineering specifically focusing on the context communication between users and agents to ease knowledge exchange and enhance user control.

\subsection{Design Implications for Controlled, Critical, and Active Deliberation with Multi-agents }
% \yiren{I found two more papers that are highly relevant but need some more time to think through how to integrate them: \cite{Yang2023BeyondTBA}\cite{Holstein2025FromCTA}}

\subsubsection{Fostering Critical Thinking through Visualizing Adversarial Discourse}
% \ta{I like this takeaway}
Participants generally valued the adversarial discourse in \systemName\, suggesting further exploration of design strategies that can foster such discourse in LLM-supported ideation systems. 
Additionally, some participants commented they would like the comments to be even more ``critical'' (RQ2). This is in stark contrast with existing chat interaction, where LLMs tend to generate responses that agree with the user's given stance (i.e., sycophancy~\cite{sharma2023towards}). Our findings suggest the value of adversarial discourse in LLM-supported ideation systems, also drawing from past research related to socio-cognitive conflict~\cite{mugny1978socio} and educational psychology~\cite{mead1994enhancing}.
% \ta{I might cut these considerations}There are three aspects to this design consideration: 1) how to design agent behaviors to encourage more critical stances; 2) how to design visualization to better convey agents' rationales and thinking processes and support users' evaluation and sensemaking; and finally, 3) how to design user interactions to better support users' steering of agents' critical thinking process.
In the context of knowledge-intensive ideation systems, adversarial discourse can be supported by designing agents that are encouraged to take on diverse and potentially conflicting perspectives, yet still grounded in evidence-based reasoning~\cite{Chang2024SocraSynthMRA,Hou2025EduThink4AITEA,Wang2023LearningTBA}.
This also covers the aspect of diversity potentially brought by adversarial deliberation, as underlined by past research suggesting that diverse perspectives can enhance collective problem-solving capabilities~\cite{page2008difference}.

Additionally, in order to support better visualization and sensemaking, design considerations can include implementing transparency features, such as temporal visualizations of an agent's reasoning paths and confidence scores for its claims and suggestions. 
These features could also help users better gauge the certainty of agent responses in terms of their expertise and the evidence gathered (e.g., literature surveyed), thus granting them stronger agency and confidence in the decision-making process~\cite{karran2022designing}.
These design considerations extend to broader educational technology contexts, where AI agents may act as facilitators of learning rather than mere information retrievers~\cite{chiang2004redesigning,garces2024takes}.

Finally, future iterations of LLM-based Multi-Agent Systems should also consider enhancing the design to better facilitate human control and agency over multi-agent collaboration to improve the collaborative output \cite{shneiderman2020human}. This aligns with the concept of adjustable autonomy, where users can dynamically manage the level of agent intervention by transferring decision-making control to the human in key situations~\cite{scerri2002adjustable}. 
Specific design improvements could include interactive mechanisms allowing users to perform direct manipulations of agents' collaboration or discussion history, and directions going forward.

\subsubsection{Balancing User Control and Agent Autonomy in Multi-Agent Ideation Systems: a Friction-Guided Approach}
Balancing between user control and system autonomy in interactive ideation and information exploration systems has been a long-standing challenge in HCI and information retrieval research~\cite{marchionini2006exploratory,Holstein2025FromCTA}. 
Wider industrial adoption also exists such as the recent OpenAI's branching feature that allows users to create multiple parallel threads of exploration from a single starting point~\cite{openai2025branchingdiscussion}, and similar branching design has also been adopted earlier in other LLM-powered systems such as Claude.
But still, in a more knowledge-intensive context, it remains a design challenge in terms of how to blend user control with quality of retrieved information~\cite{zheng2025deepresearcher,jiang2024into} --- \textit{how can system design grant user control while maintaining the depth of exploration?}

\systemName\ piloted one potential design that can effectively support the diversity of exploration while preserving the depth of discussion. This is highly relevant to the design of future knowledge-intensive search systems, such as deep research systems~\cite{huang2025deep,zheng2025deepresearcher}, where users need to balance between exploring diverse perspectives and engaging in deep discussions. 
\systemName's design can be seen as an alternative way to support a mixture of both automated and controllable exploration when it comes to ideation and knowledge synthesis, complementing existing methods used in existing research aiming to balance between amount of required user input and the amount of relevant information the system can provide in return, such as works by \citet{liu2025personaflow} and \citet{chen2025coexploreds}. 
Still, future iterations following this design direction should consider the friction-efficiency trade-off. Practically speaking, designs should consider 1) minimizing the friction when interactions encourage participants to perform active thinking; and 2) providing hybrid options that allow users to switch between low-friction and high-friction modes based on their current needs and goals.
Contextualizing this in the design of \systemName, when designing a mind map, one potential improvement is to keep exploration (interpretation) and synthesis within the same thinking space~\cite{andrews2010space} to reduce the unnecessary cognitive switching cost. To the end of balancing control and efficiency, a hybrid design that incorporates follow-up chatting in a forum-style interface could further reduce cognitive switching costs while preserving the benefits of structured deliberation.
Additionally, for existing systems that have already considered the design option of branching~\cite{openai2025branchingdiscussion}, offering an integrated or alternative visualization that offers overviews and navigation could further enhance user experience and support deeper exploration. 

Our findings also suggest that while both \systemName\ and group-chat interface are valuable, supporting complementary facets of critical thinking, the forum-style, panel-like design often included structured reasoning (e.g., organizing the motivation of proposals instead of copy-pasting), internalization of knowledge (e.g., awareness of knowledge gaps when selecting agents), and application of knowledge (e.g., seeking generalizable conclusions), whereas information-seeking was found to be more present under the traditional linear design.
For use cases where information gathering is the primary goal, less control and more hand-off to agents could be desirable, but for ideation systems where in-depth reasoning and critical thinking are the main goals, the introduced friction for stronger user control could be beneficial~\cite{ward2011productive}. 
% \yun{not very sure about the last sentence, but I feel this paragraph can moved to design implications. }

% The nested threading design enables remixing of topics which potentially leads to novel and serendipitous discoveries, which can be integrated into future designs of knowledge-intensive ideation systems~\cite{hu2024nova,radensky2024scideator}.
% The findings also surface a potential design trade-off: benefits of promoting analytical reasoning in the structured forum, and rapid information seeking and reflection in the chat modality. A hybrid design that incorporates follow-up chatting in a forum-like interface could further reduce cognitive switching costs while preserving benefits of structured deliberation.

% \yiren{The other unique insight: the original design intention for the mind-map was to support overview and navigation before the forum, but some users also prefer to use it after they read the details in forum, as a way to summarize and reflect...}
\section{Limitations and Future Work}
While the user study revealed the potential of our proposed system for supporting knowledge-intensive ideation tasks, this study has several limitations that are noteworthy. 
First, we conducted our user study within the context of interdisciplinary research. We chose this context as an example of knowledge work that requires advanced reasoning, although the generalization of insights from this specific context to broader applications needs to be further validated.
Second, we did not strictly control the participants' familiarity with the topics and research experience, as we would in a controlled study, which leaves room for potential confounds from the variance in participants' backgrounds. Future studies may consider conducting more systematic investigations of user behavior and perceptions.
Additionally, the design of \systemName\ will benefit from further contextualization in users' workflow for gathering understandings, including longer-term usage patterns, for which a field study could be in order.

\section{Conclusion}
In this paper, we conducted an exploratory within-subject study (N=18) to examine how different interaction patterns for user control over LLM-based multi-agent collaborations shape interdisciplinary research ideation. We compared two different designs with varied levels of user control: one offering features that allow users to form ad-hoc discussion panels (i.e., \systemName) and the other adopting a common single-stream group-chat interface. 
We found that \systemName\ elicited more critical-thinking activities and structured proposal revisions, whereas the group chat condition was used more for tasks with specific information-seeking. 
The findings suggest treating user control as productive friction, making agent reasoning and deliberation adversarial yet grounded, and adopting hybrid designs that balance between user control and information-seeking efficiency.  

% \input{99-appendix}

% \begin{acks}
% \end{acks}

\bibliographystyle{ACM-Reference-Format}
\bibliography{sample-base}

% \appendix
\clearpage
% \newpage
\appendix
\onecolumn
\section{Appendices}
\label{appendix}
%TC:ignore

\subsection{Pre-Session Survey}
\label{apdx:survey-items}
\subsubsection{Basic Information}
\begin{itemize}
    \item \textbf{Familiarity:} On a scale of 1 to 7, how familiar are you with the topic written in the initial proposal? (1: Not familiar at all, 7: I consider myself an expert in this domain)
    \item \textbf{Trust in GenAI:} Overall, which statement best describes your level of trust in generative AI (GenAI)?
    \begin{itemize}
        \item Very low - I rarely or never use GenAI and doubt the accuracy of its information.
        \item Low - I occasionally consult GenAI but remain skeptical of its reliability.
        \item Moderate - I sometimes use GenAI; it can be helpful, but I still double‑check its answers.
        \item High - I frequently use GenAI, find it dependable, and believe it helps me solve many problems.
        \item Very high - I actively rely on GenAI; I feel confident in its accuracy and usefulness across a wide range of tasks.
    \end{itemize}
\end{itemize}

\subsubsection{Self-Assessment Of Perceived Interdisciplinary Topic Clarity}
On a scale of 1-7 (Strongly disagree to Strongly agree):
\begin{itemize}
    \item \textbf{Conceptual Clarity:} I'm well-informed about the core concepts within the topic.
    \item \textbf{Methodological Clarity:} I know clearly the methods/approaches used within this field.
    \item \textbf{Role clarity:} I understand well who (which discipline/which colleague or expert) does what.
    \item \textbf{Communication clarity:} I feel confident explaining this topic to a mixed audience.
\end{itemize}

\subsubsection{Self-Assessment of Proposal Quality}
On a scale of 1-7 (Strongly disagree to Strongly agree):
\begin{itemize}
    \item \textbf{Coverage:} The proposal clearly explains an important research gap and demonstrates a comprehensive understanding of prior work.
    \item \textbf{Significance and Novelty:} The proposed study offers an original contribution beyond existing solutions.
    \item \textbf{Relevance:} The content in the proposal is well-aligned with the original/proposed research idea.
    \item \textbf{Depth:} The research questions, design, data-collection, and analysis plan are described in sufficient detail to convincingly answer the stated aims.
    \item \textbf{Feasibility:} The research idea proposed is feasible.
\end{itemize}

% \subsection{Participants Demographics}
% \yiren{Add table of demographics}

\subsection{Post-Survey Usability Questions and Ratings}
\subsubsection{Post-Survey Usability Questions}
\begin{itemize}
    \item \textbf{Capabilities:} The system’s capabilities meet my requirements.
    \item \textbf{Frustration:} Using the system is a frustrating experience.
    \item \textbf{Ease of Use:} I thought the system was easy to use.
    \item \textbf{Corrections:} I have to spend too much time correcting things with this system.
\end{itemize}

\subsubsection{Post-Survey Cognitive Load Questions}
\begin{itemize}
    \item \textbf{Mental Demand:} How mentally demanding was the task?
    \item \textbf{Effort:} How hard did you have to work to achieve your goal?
    \item \textbf{Stress:} How irritated, stressed, or annoyed did you feel?
\end{itemize}

\subsubsection{Post-Survey Usability and Cognitive Load Ratings}
\label{apdx:post-survey-ratings}
\begin{figure}[h]
    \centering
    \begin{subfigure}[b]{0.48\linewidth}
        \includegraphics[width=\linewidth]{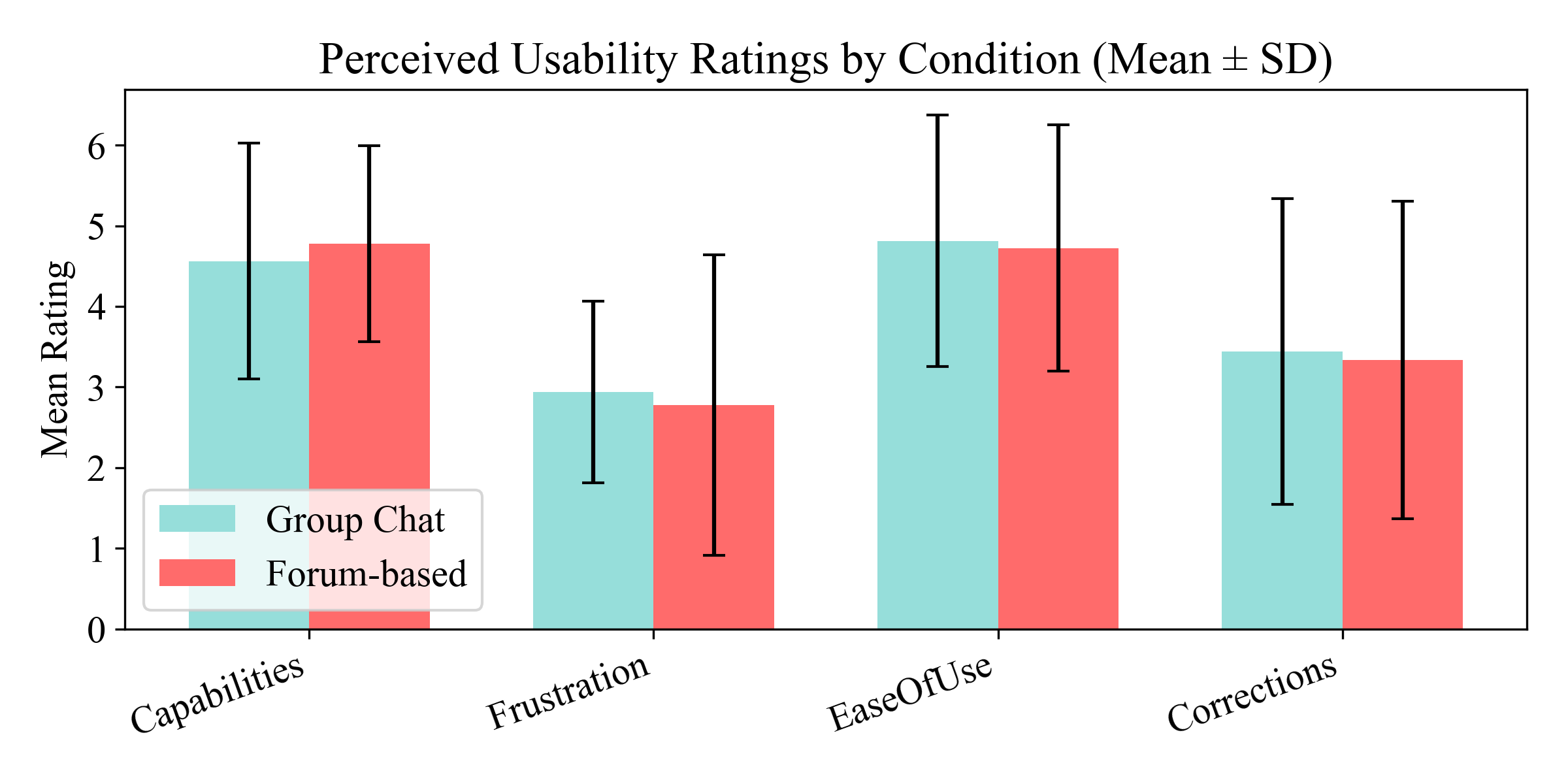}
        \caption{Usability Ratings}
        \label{fig:usability-comparison}
    \end{subfigure}\hfill
    \begin{subfigure}[b]{0.48\linewidth}
        \includegraphics[width=\linewidth]{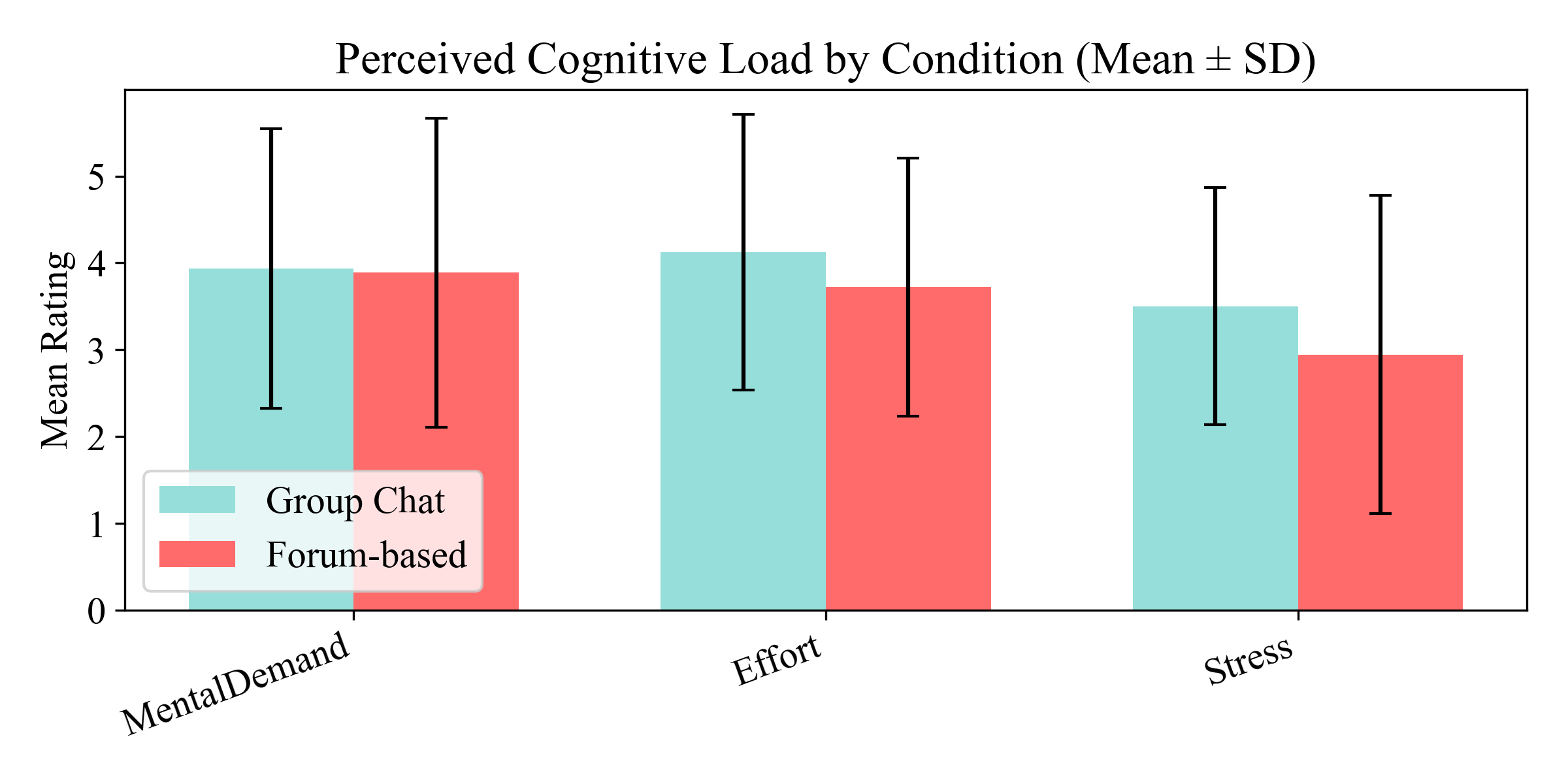}
        \caption{Cognitive Load Ratings}
        \label{fig:cognitive-comparison}
    \end{subfigure}
    \caption{Post-survey comparison of (a) usability and (b) cognitive load ratings between conditions.}
    \label{fig:usability-cognitive-combined}
    \Description{
    Figure with two bar charts (means ± SD). Left—Usability: both conditions rate capabilities and ease of use around 4.5–5; frustration about 3; corrections about 3.3. Right—Cognitive load: mental demand ≈4 in both; effort and stress are modestly higher in Group Chat than Forum-based.
    }
\end{figure}

Detailed plots of post-survey proposal quality evaluation questions results as shown in~\cref{fig:usability-cognitive-combined}.

\subsubsection{Post-Survey Proposal Quality Evaluation Questions}
\begin{itemize}
    \item \textbf{Coverage:} The proposal clearly explains an important research gap and demonstrates a comprehensive understanding of prior work.
    \item \textbf{Significance and Novelty:} The proposed study offers an original contribution beyond existing solutions.
    \item \textbf{Relevance:} The content in the proposal is well-aligned with the original/proposed research idea.
    \item \textbf{Depth:} The research questions, design, data-collection, and analysis plan are described in sufficient detail to convincingly answer the stated aims.
    \item \textbf{Feasibility:} The research idea proposed is feasible.
    \item \textbf{Clarity:} The research idea is clearly conveyed and easy to understand.
\end{itemize}

\subsection{Figure of comparison of Post-Survey Feature Ratings}
\begin{figure}[h!]
    \centering
    \includegraphics[width=0.75\linewidth]{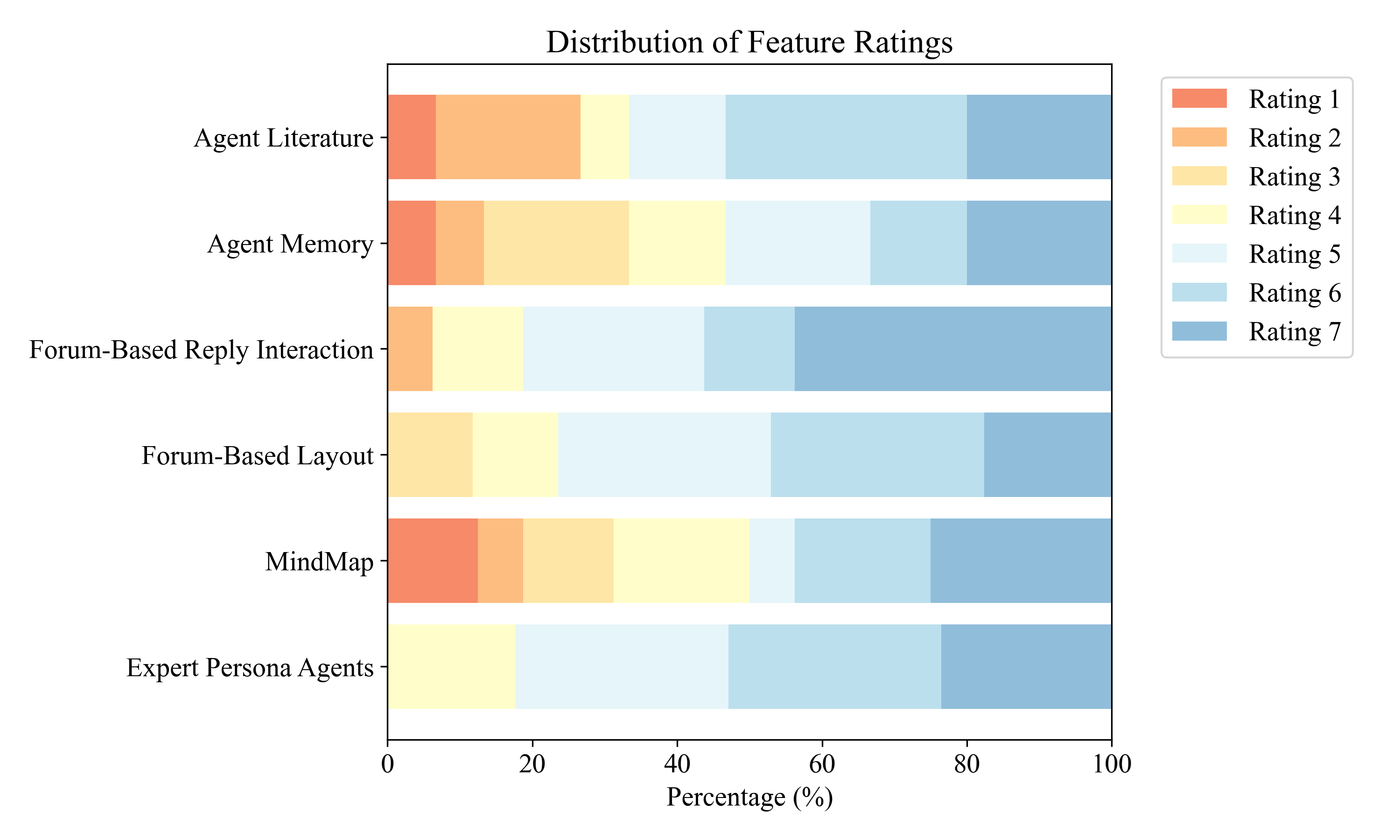}
    \caption{Overview of user ratings (overall helpfulness) for different system features (Surveys)}
    \label{fig:feature-ratings}
    \Description{
    A 100\% stacked bar chart compares six features—Expert Persona Agents, MindMap, Forum-Based Layout, Forum-Based Reply Interaction, Agent Memory, and Agent Literature—by the percentage of survey ratings from 1 (low) to 7 (high). Expert Persona Agents and Forum-Based Reply Interaction are strongly positive, with large segments in ratings 5–7. Forum-Based Layout also trends positive but with more 3–4. MindMap shows a wide spread including some 1–2 and a moderate share of 5–7. Agent Memory is mixed across 3–7 with some lows. Agent Literature concentrates in mid/low ratings (2–4) with smaller high-rating segments. X-axis shows percentage from 0–100; legend lists ratings 1–7.
    }
\end{figure}

\Cref{fig:feature-ratings} shows the comparison of post-survey feature ratings between conditions.

\subsection{Agent Profile and Memory Interface Screenshots}
\begin{figure}[h!]
    \centering
    %--- left pane -------------------------------------------------
    \begin{subfigure}[b]{0.48\linewidth}
        \includegraphics[width=\linewidth]{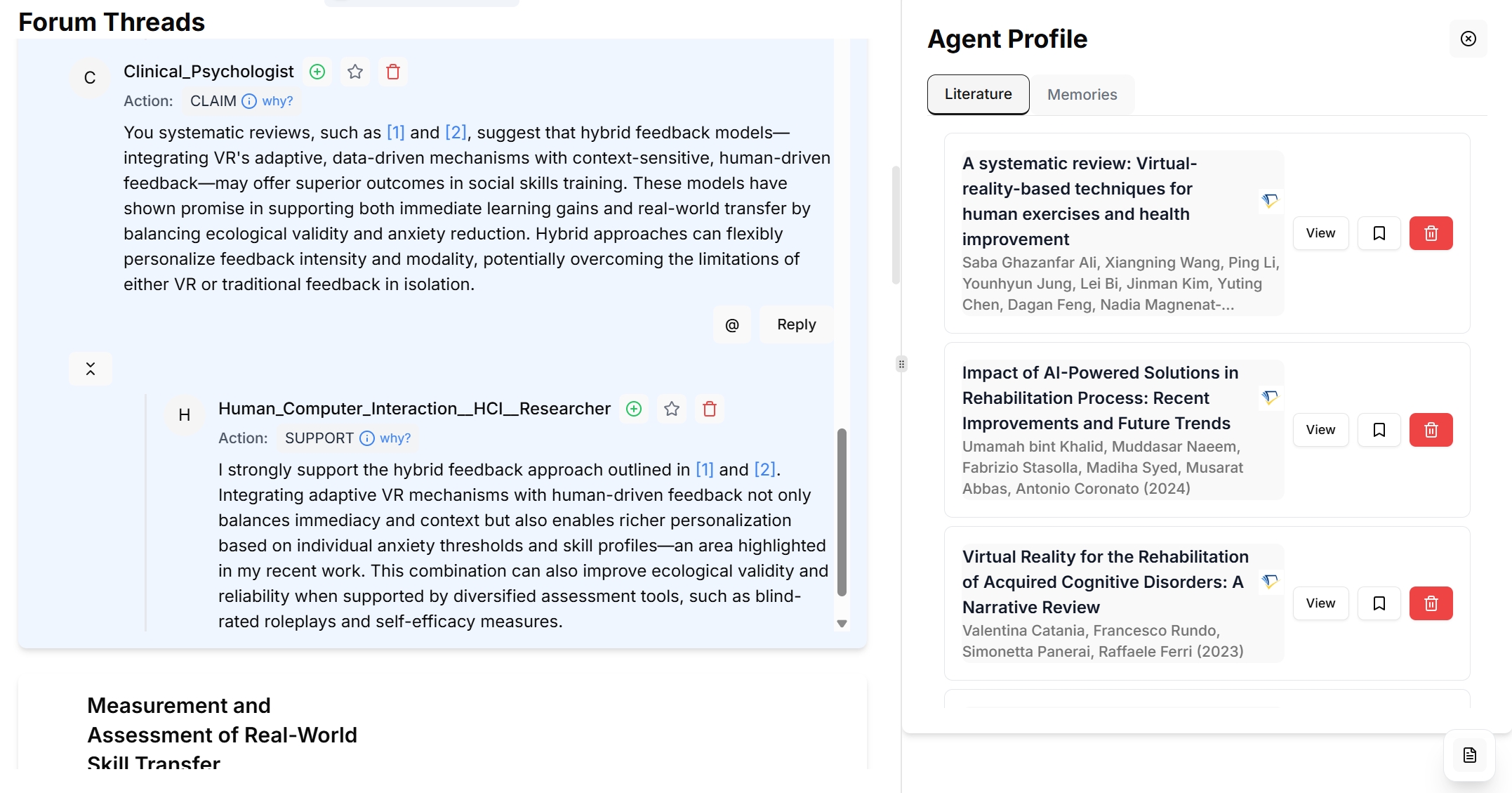}
        \caption{Agent Profile Interface}
        \label{fig:agent-profile-interface}
    \end{subfigure}\hfill
    %--- right pane ------------------------------------------------
    \begin{subfigure}[b]{0.48\linewidth}
        \includegraphics[width=\linewidth]{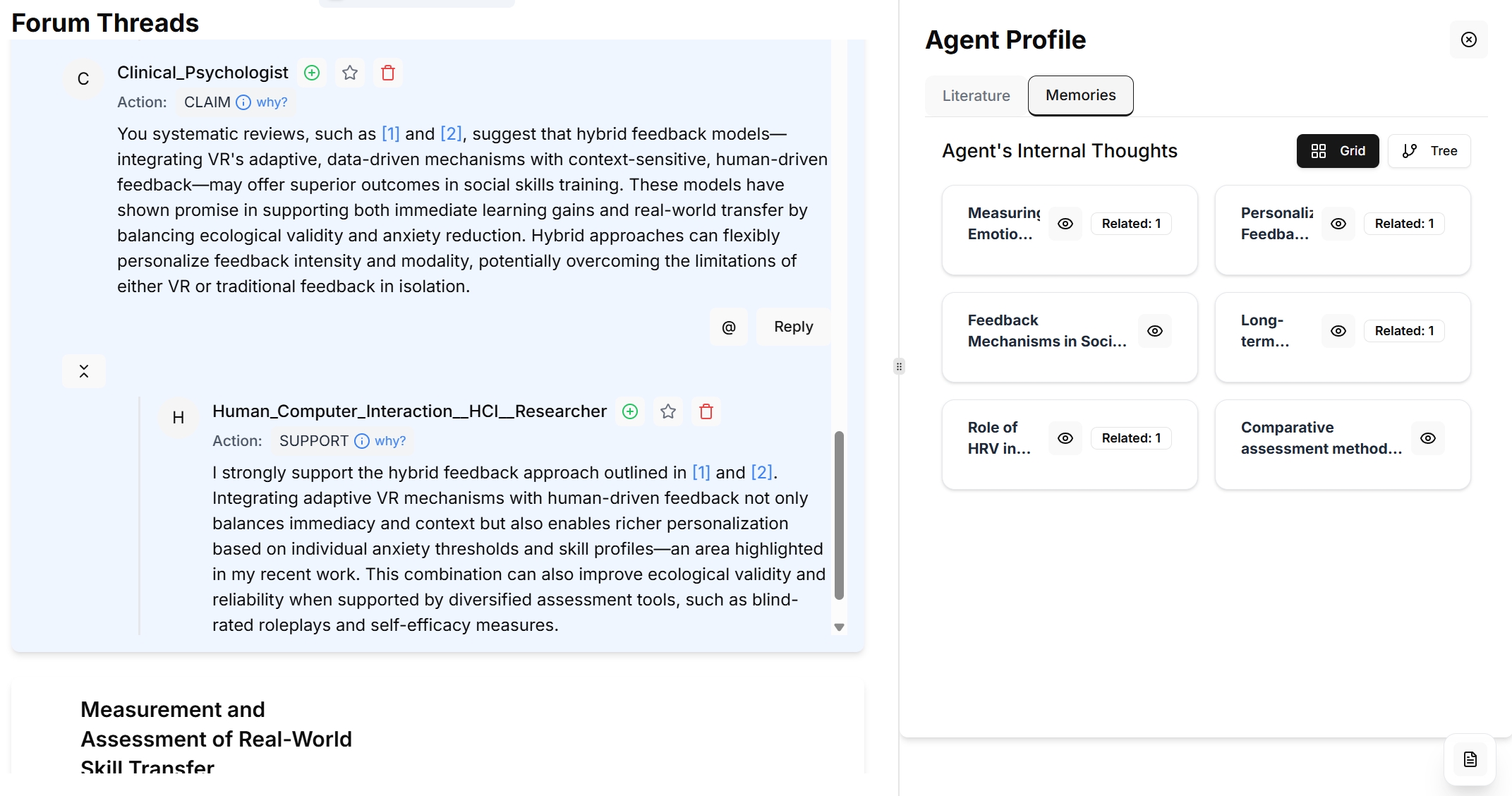}
        \caption{Agent Memory Interface}
        \label{fig:agent-memory-interface}
    \end{subfigure}

    \caption{Agent inspection interfaces: (a) profile editor and (b) memory viewer.}
    \label{fig:agent-interfaces}
    \Description{
    Side-by-side UI mockups. Left: ‘Agent Profile’ shows a literature list next to a forum thread where a Clinical Psychologist makes a claim and an HCI Researcher replies in support. Right: the same thread with the Agent Profile switched to the ‘Memories’ tab, displaying the agent’s internal thoughts as cards in a grid with ‘Related’ counts.
    }
\end{figure}

\Cref{fig:agent-interfaces} shows the screenshots of agent profile and memory inspection interfaces.

\subsection{Exit Interview Script (Semi-Structured)}
\label{apdx:exit-interview-script}
\begin{itemize}
    \item Can you walk me through your overall experience using the system to revise your proposal?
    \item Did you have any memorable moments while using the system? What were you doing right before, and what information or interaction triggered this moment?
    \item What specific information, questions, or suggestions from the agents were most helpful? Can you give an example?
    \item Did you find the different perspectives from the various personas useful? Why or why not?
    \item How do you feel about the utility/helpfulness of each feature (enumerate the feature to guide reflection)?
\end{itemize}

\subsection{Persona Taxonomy and Construction}
\label{apdx:persona-taxonomy}

\noindent\textbf{Schema.} Persona profiles following the schema below:

\begin{quote}\ttfamily
basic\_info:\\
\ \ research\_area: ...\\
\ \ short\_bio: ...\\
research\_and\_professional\_focus:\\
\ \ focus\_areas: ...\\
\ \ methodology: ...\\
\ \ publication\_channels: ...\\
skills\_and\_expertise:\\
\ \ technical\_skills: ...\\
\ \ analytical\_skills: ...\\
\ \ domain\_expertise: ...\\
personalities\_and\_characteristics:\\
\ \ communication\_style: ...\\
\ \ audience\_expertise\_level: [novice|intermediate|expert]
\end{quote}

\subsection{User Interaction Log Codebook}
\label{apdx:interaction-codebook}

\begin{table*}[h!]
\centering
\caption{Codebook for User Interaction Logs.}
\label{tab:interaction-codebook}
\begin{tabular}{p{0.1\linewidth} p{0.7\linewidth} p{0.15\linewidth}}
\toprule
\textbf{Code} & \textbf{Definition (What it captures)} & \textbf{Facione Critical-Thinking Facet} \\
\midrule
clarify & Seeks meaning of terms, concepts, or context. & Interpretation \\
\hline
expand & Requests additional detail, depth, or breadth on a point already raised. & Explanation \\
\hline
apply & Seeks concrete application of ideas, theories, or methods. & Application \& Inference \\
\hline
compare & Requests similarities, differences, or trade-offs. & Analysis \\
\hline
critique & Asks for judgment of merit, rigor, or limitations. & Evaluation \\
\hline
design & Requests creation or refinement of a study, system, or framework. & Synthesis \& Application \\
\hline
method & Seeks specific methodological choices, measurements, or analyses. & Analysis \& Application \\
\hline
data-seek & Explicit request for references, datasets, or historic evidence. & Knowledge \\
\hline
summarize & Asks to condense ideas or research into shorter form. & Explanation \\
\hline
alternative & Seeks different angles, methods, or solutions than those proposed. & Analysis \& Evaluation \\
\hline
risk & Probes potential problems, risks, or ethical concerns. & Evaluation \& Inference \\
\hline
reflect & Comments on or questions the thinking or research process itself. & Self-Regulation \\
\hline
ethics / impact & Considers moral, societal, or policy consequences. & Evaluation \& Inference \\
\bottomrule
\end{tabular}%
\Description{
Three-column table titled “Codebook for User Interaction Logs.” It lists thirteen interaction codes with their meanings and the corresponding Facione critical-thinking facet(s):  clarify—seeks meaning of terms, concepts, or context (Interpretation);  expand—requests additional detail, depth, or breadth (Explanation);  apply—seeks concrete application of ideas, theories, or methods (Application \& Inference);  compare—requests similarities, differences, or trade-offs (Analysis);  critique—asks for judgment of merit, rigor, or limitations (Evaluation);  design—requests creation or refinement of a study, system, or framework (Synthesis \& Application); method—seeks specific methodological choices, measurements, or analyses (Analysis \& Application);  data-seek—explicit request for references, datasets, or historic evidence (Knowledge);  summarize—asks to condense ideas or research (Explanation);  alternative—seeks different angles, methods, or solutions than those proposed (Analysis \& Evaluation);  risk—probes potential problems, risks, or ethical concerns (Evaluation \& Inference);  reflect—comments on or questions the thinking or research process itself (Self-Regulation);  ethics / impact—considers moral, societal, or policy consequences (Evaluation \& Inference).
}
\end{table*}

\section{T18 Proposal Edits (Full)}
\label{apdx:t18-proposal-edit-full}
It is worth noting that, T18's proposal edits during the forum condition involves stronger thinking and reasoning from themselves, whereas during the group chat the edits are mostly copy pasting~\cref{box:t18-human-robot-interaction-revision-full}.
\begin{tcolorbox}[enhanced, breakable, float=!ht, colback=white, colframe=black!40,
  label={box:t18-human-robot-interaction-revision-full}]
\footnotesize
\textbf{T18's proposal edits using the group chat baseline: Human-Robot-Interaction}

\diffctx{Motivation: }
\diffnochg{how to improve transparency of human-robot-interaction specifically in a human - multiple robots settings}

\medskip
\diffctx{Related Work: } 
\diffnochg{human robot one on one interaction but not much about robot team}
\diffadd{definition: <copied>transparency in robot-team collaboration is fundamentally about making both the processes and intentions of the robots comprehensible and visible to human team members.</copied>}

\diffadd{social perspective related work: SAT} 

\diffadd{<copied>effective explanation interfaces must support “drill-down” capability, allowing a human to start with a high-level summary and then dig into details as needed [2].</copied>}

\diffadd{<copied>Another critical challenge as teams scale is maintaining mutual awareness—robots must keep track of not only their own state but the human’s knowledge, attention, and goals, adapting their transparency accordingly. In dynamic, high-pressure scenarios like firefighting, traceability and meaningful human control need to be supported by both automated logging and live interactive explanations</copied>}

\diffadd{<copied>adaptive transparency At a sociotechnical level, transparency also involves cues like body orientation, signaling “who knows what” in the team, or highlighting moments where the robot’s autonomy level is shifting. These are often not just about explanation content, but also timing, modality (verbal, visual, gestural), and invitation for user interaction—core aspects identified in both SAT and DARPA XAI outputs [2] [3]. I’d argue that future research should explore co-adaptive transparency, where both robots and humans shape the level and type of explanatory interaction together, influenced by ongoing context—a dynamic still underexplored in both interface design and team protocols. What do others see as necessary methodological advances to capture and refine this adaptivity in multi-robot teamwork?</copied>}

\medskip
\diffctx{Methods: }
\diffnochg{literature rievew, tool design, propotype development, evaluation}

\diffadd{
<copied>system design: expose their internal decision-making processes  (added point: transparency hinges not just on exposing internal state or logs, but on providing explanations that are actionable and adaptable to the context and user expertise)
adapt their communication style to different users, 
humans should also be able to probe, question, and receive feedback in ways that make sense in context. </copied>
}

\diffadd{
ideas for overcoming the trade-off between detail and overload?
}

\diffadd{
<copied>senario design: From the qualitative side, I’d suggest deploying adaptive transparency in a hospital logistics context—imagine autonomous robots delivering sensitive supplies, where nurses and staff have varying expertise and urgency levels. Transparency would flex: routine deliveries need minimal prompts, while high-priority or unexpected events (e.g., medication reroute due to emergency) trigger detailed, user-tailored explanation (visual dashboard for seasoned staff; stepwise verbal walk-through for new users). Your user studies could use ethnographic observation and post-task interviews to explore not just trust and error rates, but also deeper aspects of social acceptability and staff perceptions of agency and accountability—key factors in organizational buy-in, as highlighted in sociotechnical and XAI deployments[1]. This offers rich ground for multi-method experiments! Curious what others think about capturing these “soft” social outcomes alongside quantitative measures.</copied>
}

\diffadd{
<copied>evaluation metrics: That’s a fantastic take—I’d echo the critical value of “soft” outcome measures, especially in settings like hospitals where team cohesion and perceived accountability heavily influence deployment success. Ethnographic and mixed-methods approaches could reveal nuanced effects of adaptive transparency—for example, whether more granular explanations at critical moments foster not just trust, but also long-term user empowerment and error mitigation, as seen in XAI user studies[1][2]. For experimentation, you could instrument both quantitative metrics (task success, workload, response latency) and qualitative feedback (semi-structured interviews probing perceived clarity, fairness, team fit). This hybrid strategy is well-supported by XAI program guidelines and ensures the adaptivity isn’t just technically sound, but genuinely aligns with user values and organizational culture. I’d encourage prototyping transparency “knobs” with multi-modal options (visual + verbal) and field-staging them—adaptive transparency is as much about fit to workflow and norm as explanation content. </copied>
}

\medskip
\diffctx{Potential Outcomes: }
\diffnochg{more efforts should be put on how to show the coordination between the robot teammates, and level of this transparency should be carefully considered}

% \tcblower
% Differences between T16's initial and revised proposal using \systemName.
\end{tcolorbox}

%TC:endignore

\end{document}